\documentclass{emulateapj}

\usepackage{times}
\usepackage{graphicx}
\usepackage{amssymb}
\usepackage{amsmath}
\usepackage{pifont}
\usepackage{graphics}
\usepackage{xcolor}
\usepackage{epsfig}
\usepackage{verbatim}
\usepackage{booktabs}

\newcommand{\bjdtdb}{BJD$_\textrm{TDB}\ $}      
\newcommand{\feh}{\ensuremath{\left[{\rm Fe}/{\rm H}\right]}}  
\newcommand{\meh}{\ensuremath{\left[{\rm m}/{\rm H}\right]}} 
\newcommand{\ecosw}{\ensuremath{e\cos{\omega_\star}}} 
\newcommand{\esinw}{\ensuremath{e\sin{\omega_\star}}}
\newcommand{\teff}{\ensuremath{T_{\rm eff}}}
\newcommand{\msun}{\ensuremath{\,M_\Sun}}
\newcommand{\rsun}{\ensuremath{\,R_\Sun}}
\newcommand{\lsun}{\ensuremath{\,L_\Sun}}
\newcommand{\mj}{\ensuremath{\,M_{\rm J}}}
\newcommand{\rj}{\ensuremath{\,R_{\rm J}}}
\newcommand{\fave}{\langle F \rangle}
\newcommand{\fluxcgs}{10$^9$ erg s$^{-1}$ cm$^{-2}$}

\newcommand{\kms}{$\,$km$\,$s$^{-1}$}
\newcommand{\ms}{$\,$m$\,$s$^{-1}$}

\begin{document}
\title{KELT-16\MakeLowercase{b}: A highly irradiated, ultra-short period hot Jupiter nearing tidal disruption}

\author{
Thomas E. Oberst$^{1}$,
Joseph E. Rodriguez$^{2}$,
Knicole D. Col\'on$^{3,4}$,
Daniel  Angerhausen$^{5,6}$,
Allyson  Bieryla$^{7}$,
Henry  Ngo$^{8}$,
Daniel J. Stevens$^{9}$,
Keivan G. Stassun$^{2,10}$,
B. Scott  Gaudi$^{9}$,
Joshua  Pepper$^{11}$,
Kaloyan  Penev$^{12}$,
Dimitri  Mawet$^{13,14}$,
David W. Latham$^{7}$,
Tyler M. Heintz$^{1}$,
Baffour W. Osei$^{10}$,
Karen A. Collins$^{2,10}$,
John F. Kielkopf$^{15}$,
Tiffany  Visgaitis$^{16}$,
Phillip A. Reed$^{16}$,
Alejandra  Escamilla$^{17}$,
Sormeh  Yazdi$^{17}$,
Kim K. McLeod$^{17}$,
Leanne T. Lunsford$^{18}$,
Michelle  Spencer$^{18}$,
Michael D. Joner$^{18}$,
Joao  Gregorio$^{19}$,
Clement  Gaillard$^{18}$,
Kyle  Matt$^{18}$,
Mary Thea Dumont$^{20,18}$,
Denise C. Stephens$^{18}$,
David H. Cohen$^{21}$,
Eric L. N. Jensen$^{21}$,
Sebastiano Calchi Novati$^{22,23}$,
Valerio  Bozza$^{23,24}$,
Jonathan  Labadie-Bartz$^{11}$,
Robert J. Siverd$^{25}$,
Michael B. Lund$^{2}$,
Thomas G. Beatty$^{26,27}$,
Jason D. Eastman$^{7}$,
Matthew T. Penny$^{9}$,
Mark  Manner$^{28}$,
Roberto  Zambelli$^{29}$,
Benjamin J. Fulton$^{30,31}$,
Christopher  Stockdale$^{32}$,
D. L. DePoy$^{33,34}$,
Jennifer L. Marshall$^{33,34}$,
Richard W. Pogge$^{9}$,
Andrew  Gould$^{9}$,
Mark  Trueblood$^{35}$,
Patricia  Trueblood$^{35}$
}

\affil{}
\affil{$^{1}$Department of Physics, Westminster College, New Wilmington, PA 16172, USA}
\affil{$^{2}$Department of Physics and Astronomy, Vanderbilt University, Nashville, TN 37235, USA}
\affil{$^{3}$NASA Ames Research Center, M/S 244-30, Moffett Field, CA 94035, USA}
\affil{$^{4}$Bay Area Environmental Research Institute, Petaluma, CA 94952, USA}
\affil{$^{5}$NASA Goddard Space Flight Center, Exoplanets and Stellar Astrophysics Laboratory, Code 667, Greenbelt, MD 20771, USA}
\affil{$^{6}$CSH Fellow, Center for Space and Habitability, Universit\"at Bern, Sidlerstrasse 5, 3012 Bern, Switzerland}
\affil{$^{7}$Harvard-Smithsonian Center for Astrophysics, Cambridge, MA 02138, USA}
\affil{$^{8}$Division of Geological and Planetary Sciences, California Institute of Technology, Pasadena, CA 91125, USA}
\affil{$^{9}$Department of Astronomy, The Ohio State University, Columbus, OH 43210, USA}
\affil{$^{10}$Department of Physics, Fisk University, Nashville, TN 37208, USA}
\affil{$^{11}$Department of Physics, Lehigh University, Bethlehem, PA 18015, USA}
\affil{$^{12}$Department of Astrophysical Sciences, Princeton University, Princeton, NJ 08544, USA}
\affil{$^{13}$Department of Astronomy, California Institute of Technology, Pasadena, CA 91125, USA}
\affil{$^{14}$Jet Propulsion Laboratory, California Institute of Technology, Pasadena, CA 91109, USA}
\affil{$^{15}$Department of Physics and Astronomy, University of Louisville, Louisville, KY 40292, USA}
\affil{$^{16}$Department of Physical Sciences, Kutztown University, Kutztown, PA 19530, USA}
\affil{$^{17}$Department of Astronomy, Wellesley College, Wellesley, MA 02481, USA}
\affil{$^{18}$Department of Physics and Astronomy, Brigham Young University, Provo, UT 84602, USA}
\affil{$^{19}$Atalaia Group and CROW Observatory, Portalegre, Portugal}
\affil{$^{20}$Department of Astronomy and Astrophysics, University of California Santa Cruz, Santa Cruz, CA 95064, USA}
\affil{$^{21}$Department of Physics and Astronomy, Swarthmore College, Swarthmore, PA 19081, USA}
\affil{$^{22}$Infrared Processing and Analysis Center and NASA Exoplanet Science Institute, California Institute of Technology, Pasadena, CA 91125, USA}
\affil{$^{23}$Dipartimento di Fisica ``E. R. Caianiello,'' Universit\`a di Salerno, I-84084 Fisciano (SA), Italy}
\affil{$^{24}$Istituto Nazionale di Fisica Nucleare, Sezione di Napoli, 80126 Napoli, Italy}
\affil{$^{25}$Las Cumbres Observatory Global Telescope Network, Santa Barbara, CA 93117, USA}
\affil{$^{26}$Department of Astronomy and Astrophysics, The Pennsylvania State University, University Park, PA 16802, USA}
\affil{$^{27}$Center for Exoplanets and Habitable Worlds, The Pennsylvania State University, University Park, PA 16802, USA}
\affil{$^{28}$Spot Observatory, Nashville, TN 37206, USA}
\affil{$^{29}$Societ\`a Astronomica Lunae, Castelnuovo Magra 19030, Italy}
\affil{$^{30}$Institute for Astronomy, University of Hawaii, Honolulu, HI 96822, USA}
\affil{$^{31}$NSF Graduate Research Fellow}
\affil{$^{32}$Hazelwood Observatory, Churchill, Victoria, Australia}
\affil{$^{33}$George P. and Cynthia Woods Mitchell Institute for Fundamental Physics and Astronomy, Texas A \& M University, College Station, TX 77843, USA}
\affil{$^{34}$Department of Physics and Astronomy, Texas A \& M University, College Station, TX 77843, USA}
\affil{$^{35}$Winer Observatory, Sonoita, AZ 85637, USA}

\shorttitle{KELT-16\MakeLowercase{b}}

\begin{abstract}

We announce the discovery of KELT-16b, a highly irradiated, ultra-short period hot Jupiter transiting the relatively bright ($V = 11.7$) star TYC 2688-1839-1/KELT-16. A global analysis of the system shows KELT-16 to be an F7V star with $T_\textrm{eff} = 6236\pm54$ K, $\log{g_\star} = 4.253_{-0.036}^{+0.031}$, $\feh = -0.002_{-0.085}^{+0.086}$, $M_\star = 1.211_{-0.046}^{+0.043} M_\odot$, and $R_\star = 1.360_{-0.053}^{+0.064} R_\odot$. The planet is a relatively high mass inflated gas giant with $M_\textrm{P} = 2.75_{-0.15}^{+0.16} M_\textrm{J}$, $R_\textrm{P} = 1.415_{-0.067}^{+0.084} R_\textrm{J}$, density $\rho_\textrm{P} = 1.20\pm0.18$ g cm$^{-3}$, surface gravity $\log{g_\textrm{P}} = 3.530_{-0.049}^{+0.042}$, and $T_\textrm{eq} = 2453_{-47}^{+55}$ K. The best-fitting linear ephemeris is $T_\textrm{C} = 2457247.24791\pm0.00019$ \bjdtdb and $P = 0.9689951 \pm 0.0000024$ d. KELT-16b joins WASP-18b, -19b, -43b, -103b, and HATS-18b as the only giant transiting planets with $P < 1$ day. Its ultra-short period and high irradiation make it a benchmark target for atmospheric studies by HST, Spitzer, and eventually JWST. For example, as a hotter, higher mass analog of WASP-43b, KELT-16b may feature an atmospheric temperature-pressure inversion and day-to-night temperature swing extreme enough for TiO to rain out at the terminator. KELT-16b could also join WASP-43b in extending tests of the observed mass-metallicity relation of the Solar System gas giants to higher masses. KELT-16b currently orbits at a mere $\sim$ 1.7 Roche radii from its host star, and could be tidally disrupted in as little as a few $\times 10^{5}$ years (for a stellar tidal quality factor of $Q_*' = 10^5$). Finally, the likely existence of a widely separated bound stellar companion in the KELT-16 system makes it possible that Kozai-Lidov oscillations played a role in driving KELT-16b inward to its current precarious orbit.
\end{abstract}


\maketitle

\section{Introduction}
\label{sec:Intro}

The detection of the first transiting exoplanet, HD 209458b, in 1999 \citep{Charbonneau:2000,Henry:2000} helped to propel and inspire small aperture ground-based synoptic searches for transits such as the Trans-Atlantic Exoplanet Search (TrES; \citealt{Alonso:2004}), XO \citep{McCullough:2005}, the Wide Angle Search for Planets (WASP; \citealt{Pollacco:2006}), the Hungarian-made Automated Telescope Network (HATNet; \citealt{Bakos:2004}), and others -- including the Kilodegree Extremely Little Telescope (KELT; \citealt{Pepper:2007, Pepper:2012}). Over the decade-and-a-half following that first transit detection, these surveys have collectively discovered and characterized $\sim 250$ transiting exoplanets, nearly all of which are gas giant planets ($0.1 \lesssim M \lesssim 13 M_\textrm{J}$) in short period ($P \lesssim 10$ d) orbits,\footnote{The Extrasolar Planets Encyclopedia \url{http://www.exoplanet.eu}, accessed on July 15, 2016.}  or so-called ``hot Jupiters,'' since their deeper and more frequent transits help to overcome the noise and phase coverage limitations of ground-based observations. 

Nearly 10 times more exoplanets and a wider range of exoplanet types have been discovered and studied by the Kepler space telescope,\footnotemark[\value{footnote}] launched in 2009 and still active \citep{Borucki:2010, Howell:2014, Coughlin:2016}. Nevertheless, because most Kepler systems are too faint or have orbital periods too long for detailed follow-up observations by existing instruments, transiting hot Jupiters discovered by ground-based surveys remain among the most valuable targets for exoplanet science. And because of Kepler's limited sky coverage, small ground-based telescope networks such as KELT continue to play a critical role in discovering and characterizing these planets, and in identifying the most promising hot Jupiters for follow-up observations by the Hubble Space Telescope (HST), Spitzer Space Telescope, and, eventually, the James Webb Space Telescope (JWST) -- a trend likely to continue until the launch of the Transiting Exoplanet Survey Satellite (TESS; \citealt{Ricker:2015}).

Hot Jupiters pose several important science questions. First, their formation and migration pathways are not well understood. Giant planets are thought to form by either accretion of gas onto a $\sim$ 10 $M_\Earth$ solid core within 1 - 10 AU of the host star (``core accretion''; \citealt{Pollack:1996, Lissauer:2007}) or rapid collapse due to  gravitational instability a few tens of AU from the host star (``disk instability''; \citealt{Boss:2000, Boley:2009}). Both scenarios require subsequent inward migration of the planet, which may be achieved via the planet's interaction with the protoplanetary disk (specifically Type II migration for giant planets; e.g., \citealt{DAngelo:2008}), or later via gravitational interactions with other massive planets or stars in the system (``gravitational scattering''; e.g., \citealt{Rasio:1996, Fabrycky:2007}). Attempts to observationally delineate between these possible formation and migration mechanisms have been inconclusive, necessitating the continued observation and study of hot Jupiter dynamics (e.g. \citealt{Madhusudhan:2014}). 

As these giant planets migrate closer to their star, irradiation becomes a dominant driver of hot Jupiters' physics and evolution. For instance, hot Jupiter radii seem to increase with increased stellar irradiation \citep{Demory:2011} even though 
Jupiter-mass objects are expected to have radii that are only a few tens of percent larger than Jupiter itself even when highly irradiated. This so-called ``hot Jupiter inflation problem'' has been recognized for a long time \citep{Baraffe:2003}. Proposed explanations include tidal heating, deposition of heat through the atmosphere via vertical mixing, and Ohmic dissipation (e.g., \citealt{Miller:2009, Leconte:2010, Perna:2012, Spiegel:2013, Ginzburg:2015, Ginzburg:2016} and references therein).

Transiting hot Jupiters are especially amenable to atmospheric characterization, the latest frontier in exoplanet science. During transit, transmission spectroscopy of the planet's backlit atmosphere can probe the atmospheric composition, rotation rate, wind speeds, and the presence of clouds and hazes (e.g., \citealt{Charbonneau:2002, Brogi:2016, Sing:2016}). During secondary eclipse, photometry can constrain the planet's albedo or brightness temperature while spectrally-resolved observations can probe the vertical structure of the atmosphere (e.g., \citealt{Deming:2006, Beatty:2014, Zhou:2015}). And, finally, throughout its orbit, a planet's infrared thermal emission can be monitored to construct a ``phase curve'' to study temperature profiles, day-to-night energy transport, and winds (e.g., \citealt{Knutson:2012, Zellem:2014, Stevenson:2014, Angerhausen:2015}). 

Here we report the discovery of KELT-16b, one of only six transiting giant exoplanets with $P < 1$ d. Due to its short period, extreme irradiation, and relatively bright ($V = 11.7$) host, KELT-16b presents the opportunity for both more convenient and higher signal-to-noise ratio (SNR) follow-up observations than most other hot Juipters. It is a hotter, higher mass analog of the planet WASP-43b, which has become one of the best-studied planets and is the benchmark for ultra-short period highly irradiated giants \citep{Hellier:2011, Wang:2013, Czesla:2013, Blecic:2014, Chen:2014, Murgas:2014, Kreidberg:2014, Zhou:2014}. We thus expect KELT-16b to become a similarly valuable target for HST, Spitzer, the soon-to-be-launched JWST, and other observatories in the study of exoplanet formation, migration, and atmospheric processes. 

We present the discovery and follow-up observations of KELT-16b in \S\ref{sec:Obs}, analysis in \S\ref{sec:Analysis}, false-positive scenarios in \S\ref{sec:False-Positives}, and discussion in \S\ref{sec:Discussion}.

\section{Discovery and Follow-Up Observations}
\label{sec:Obs}

\subsection{Discovery}
\label{sec:Discovery}

The Kilodegree-Extremely Little Telescope (KELT) is an all-sky photometric survey for planets transiting bright hosts. It was originally optimized to target stars of brightnesses $8 < V < 10$ -- filling a niche between the faintness limit of most RV surveys and the saturation limit of most transit surveys -- but, counting the current discovery, it has discovered planets around stars as faint as $V \sim 12$. The survey uses two telescopes, KELT-North in Sonoita, Arizona, and KELT-South in Sutherland, South Africa. Each telescope has a 26$^{\circ}\times$26$^{\circ}$ field of view and a 23$\arcsec$ pixel scale. Together these twin telescopes observe over 70\% of the entire sky with 10-20 minute cadence and $\sim$ 1 \% photometric noise (for $V \lesssim 12$; \citealt{Pepper:2007, Pepper:2012}). The survey has discovered and published 13 transiting planets around bright stars (\citealt{Siverd:2012, Beatty:2012, Pepper:2013, Eastman:2016, Collins:2014, Bieryla:2015, Fulton:2015, Kuhn:2016, Pepper:2016, Rodriguez:2016, Zhou:2016}, and the current work).

\begin{figure}
\centering 
\includegraphics[width=1\columnwidth,angle=0]{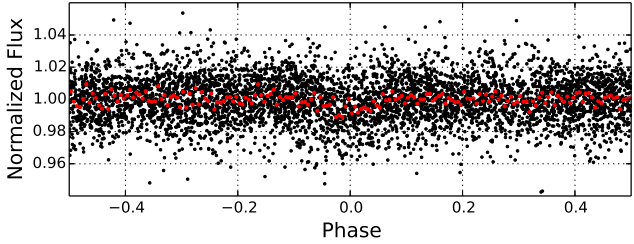}
\caption{\footnotesize KELT-16b discovery light curve. The light curve contains 5,626 observations by the KELT-North telescope over six years, phase-folded to the orbital period of 0.9690039 days from the original BLS fit. The red points represent the same data binned at $\sim$ 8 minute intervals after phase-folding.}
\label{fig:DiscoveryLC}
\end{figure}

KELT-16 is located in KELT-North survey field 12, centered on ($\alpha=21^h22^m52\fs8$, $\delta = +31\arcdeg39\arcmin56\farcs2$; J2000). We monitored this field from 2007 June to 2013 June, collecting a total of 5,626 observations. 
One of the candidates in field 12 that passed our selection cuts was matched to TYC 2688-1839-1, located at ($\alpha = 20^h57^m04\fs435, \delta= +31\arcdeg39\arcmin39\farcs57$; J2000). In the KELT-North light curve of this candidate, a significant Box-Least-Squares (BLS; \citealt{Kovacs:2002}) signal was found at a period of $P \approx 0.9690039$ d with a transit depth of $\delta \approx 6.0$ millimagnitudes (mmag) and duration of approximately 2.26 hours (our image reduction and light curve processing is described in detail in \citet{Siverd:2012}). This prompted follow-up observations of the target (\S\ref{sec:Photom}) and its eventual designation as KELT-16. The discovery light curve is shown in Figure \ref{fig:DiscoveryLC}. Properties of the host star are listed in Table \ref{tab:LitProps}. 

We note that this KELT candidate was initially incorrectly matched to TYC 2688-1883-1, located at ($\alpha = 20^h57^m01\fs59, \delta= +31\arcdeg39\arcmin37\farcs74$, J2000). It was not until the first follow-up observation (\S \ref{sec:Photom}) that it was discovered that the transit event was actually in the neighboring star TYC 2688-1839-1. This erroneous identification was a casualty of the catalog matching process, which is performed independently for the east and west (pre- and post-meridian flip) KELT data \citep{Siverd:2012}. For each object identified during point-spread function (PSF) photometry, we query the Tycho-2 catalog for stars within a $6\arcsec$ radius. The east KELT object lies $22\farcs92$ from TYC 2688-1883-1 and $23\farcs18$ from TYC 2688-1839-1, while the corresponding west KELT object lies $18\farcs18$ from TYC 2688-1883-1 and $23\farcs59$ from TYC 2688-1839-1. As a result, both objects matched to the closer TYC 2688-1883-1, and thus to each other. Errors like this underscore the challenges of matching catalogs generated from high resolution data to lower quality images, and the value of the KELT Follow-up Network (\S \ref{sec:Photom}). 

\begin{table}
\footnotesize
\centering
\caption{Stellar Properties of KELT-16.}
\begin{tabular}{llcc}
   \hline
  \hline
  Parameter & Description & Value & Ref. \\
Names\dotfill & \multicolumn{3}{l}{Tycho ID \ldots\ldots\ldots\ldots\ldots \hspace{5pt} TYC 2688-1839-1}				\\
	  & \multicolumn{3}{l}{2MASS ID \ldots\ldots\ldots \hspace{0pt} 2MASS J20570443+3139397}	\\
$\alpha_\textrm{J2000}$\dotfill	&Right Ascension (RA)\dotfill & 20:57:04.435			& 1	\\
$\delta_\textrm{J2000}$\dotfill	&Declination (Dec)\dotfill & +31:39:39.57			& 1	\\
$NUV$\dotfill           & GALEX $NUV$ mag.\dotfill & 16.14 $\pm$ 0.26 & 2 \\
$B_\textrm{~T}$\dotfill			&Tycho $B_\textrm{~T}$ mag.\dotfill & 12.28 $\pm$ 0.15		& 1	\\
$V_\textrm{~T}$\dotfill			&Tycho $V_\textrm{~T}$ mag.\dotfill & 11.72 $\pm$ 0.12		& 1	\\
$B$\dotfill		& APASS Johnson $B$ mag.\dotfill	& 12.247 $\pm$	0.062		& 3	\\
$V$\dotfill		& APASS Johnson $V$ mag.\dotfill	& 11.898 $\pm$	0.030		& 3	\\
$g'$\dotfill		& APASS Sloan $g'$ mag.\dotfill	& 11.873 $\pm$	0.103		& 3	\\
$r'$\dotfill		& APASS Sloan $r'$ mag.\dotfill	& 	11.801	$\pm$ 0.030	& 3	\\
$i'$\dotfill		& APASS Sloan $i'$ mag.\dotfill	& 	11.685 $\pm$	0.030	& 3	\\
$J$\dotfill			& 2MASS $J$ mag.\dotfill & 10.928 $\pm$ 0.023		& 4, 5	\\
$H$\dotfill			& 2MASS $H$ mag.\dotfill & 10.692 $\pm$ 0.022	& 4, 5	\\
$K_\textrm{S}$\dotfill			& 2MASS $K_\textrm{S}$ mag.\dotfill & 10.642 $\pm$ 0.016		& 4, 5	\\
\textit{WISE1}\dotfill		& \textit{WISE1} mag.\dotfill & 10.568 $\pm$ 0.023		& 6, 7	\\
\textit{WISE2}\dotfill		& \textit{WISE2} mag.\dotfill & 10.564 $\pm$ 0.020		& 6, 7 \\
\textit{WISE3}\dotfill		& \textit{WISE3} mag.\dotfill & 10.352 $\pm$ 0.053		& 6, 7	\\
\textit{WISE4}\dotfill		& \textit{WISE4} mag.\dotfill & $\le 8.787$		& 6, 7	\\
$\mu_{\alpha}$\dotfill		& NOMAD proper motion\dotfill & 5.1 $\pm$ 0.7	& 8 \\
                    & \hspace{3pt} in RA (mas yr$^{-1}$)	& & \\
$\mu_{\delta}$\dotfill		& NOMAD proper motion\dotfill 	&  0.4 $\pm$ 0.7 &  8 \\
                    & \hspace{3pt} in DEC (mas yr$^{-1}$) & & \\
$RV$\dotfill & Absolute radial \hspace{9pt}\dotfill  & $-29.7 \pm 0.1$ & This work \\
     & \hspace{3pt} velocity (\kms)  & & \\
$v\sin{i_\star}$\dotfill &  Stellar rotational \hspace{7pt}\dotfill &  7.6 $\pm$ 0.5 & This work \\
                 & \hspace{3pt} velocity (\kms)  & & \\
Spec. Type\dotfill & Spectral Type\dotfill & F7V & This work \\
Age\dotfill & Age (Gyr)\dotfill & $3.1 \pm 0.3$ & This work \\
$d$\dotfill & Distance (pc)\dotfill & $399 \pm 19$ & This work \\
$A_V$\dotfill & Visual extinction (mag.)\dotfill & $0.04 \pm 0.04$ & This work \\
$U^{*}$\dotfill & Space motion (\kms)\dotfill & $-5.4 \pm 1.3$  & This work \\
$V$\dotfill       & Space motion (\kms)\dotfill & $-13.5 \pm 0.4$ & This work \\
$W$\dotfill       & Space motion (\kms)\dotfill & $4.1 \pm 1.3$   & This work \\
\hline
\hline
\end{tabular}
\begin{flushleft} 
 \footnotesize{ \textbf{\textsc{NOTES:}}
$^{*}U$ is positive in the direction of the Galactic Center. 
    References are: $^1$\citet{Hog:2000}, $^2$\citet{Bianch:2011}, $^3$\citet{Henden:2015}, $^4$\citet{Cutri:2003}, $^5$\citet{Skrutskie:2006}, $^6$\citet{Wright:2010}, $^7$\citet{Cutri:2014}, and $^8$\citet{Zacharias:2004}.
}
\end{flushleft}
\label{tab:LitProps}
\end{table}

\subsection{Photometric Follow-up}
\label{sec:Photom}

We obtained follow-up time-series photometry of KELT-16 to verify the planet detection, check for false-positives, and better resolve the transit profile. These observations employed the KELT-North Follow-up Network (FUN), a closely-knit international collaboration of approximately 20 privately and publicly funded observatories with $\sim$ 0.25 to 1 m aperture telescopes. The member observatories span a wide range of longitudes and latitudes to maximize the temporal and sky coverage of transit events. Scientific and logistical coordination within the network is made possible by several custom software tools including the TAPIR package \citep{Jensen:2013}, which is used to predict transit events.

\begin{figure}
\vspace{.0in}
\includegraphics[width=1\linewidth]{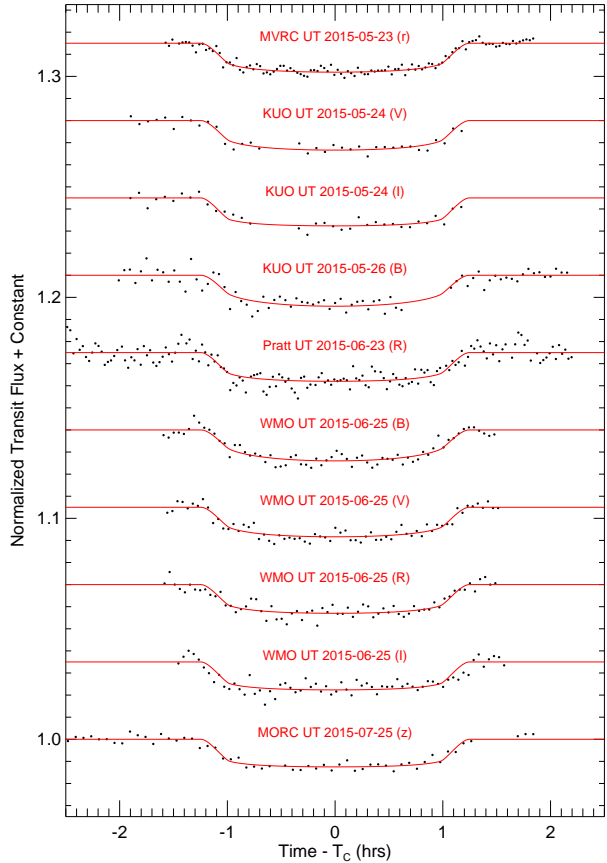}
\caption{\footnotesize Individual follow-up light curves of KELT-16b from KELT-FUN (black points) with the best fit global model (\S\ref{sec:GlobalFit}) overplotted on each curve (solid red line). The light curves are found to be achromatic and the model shows the average limb darkening weighted by the number of transits in each band. (\textit{Continued in Fig. \ref{fig:LCs1b}}.) }
\label{fig:LCs1a} 
\end{figure}

\begin{figure}
\vspace{.0in}
\includegraphics[width=1\linewidth]{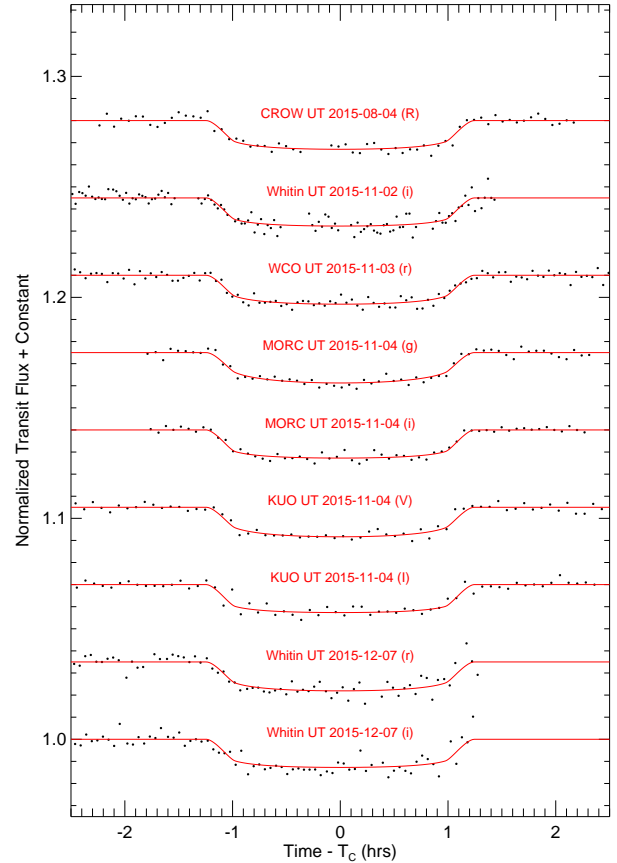}
\caption{\footnotesize Individual follow-up light curves of KELT-16b from KELT-FUN (\textit{continued from Fig. \ref{fig:LCs1a}}). }
\label{fig:LCs1b} 
\end{figure}

\begin{figure}
\vspace{0in}
\includegraphics[width=1\linewidth]{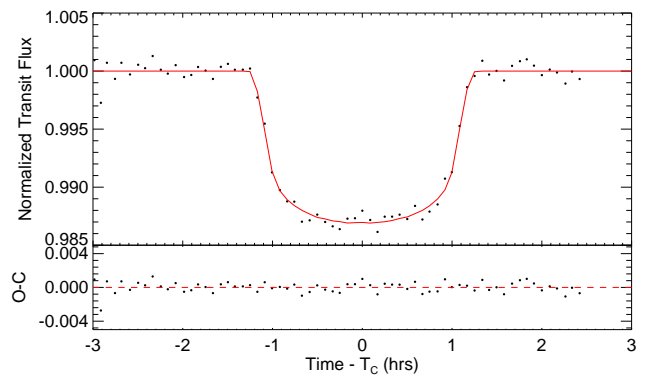}
\caption{\footnotesize All follow-up transits combined and binned in 5 minute intervals (black points) and overplotted with the best fit global model (solid red line). The model shows the average limb darkening weighted by the number of transits in each band. The differences between the data and model, or observed-corrected (O-C) residuals, are shown in the bottom panel. These binned light curve data are not used in the analysis and are presented here purely to showcase the overall statistical power of the follow-up photometry. }
\label{fig:LCs2} 
\end{figure}

KELT-FUN obtained 19 full
transit observations in the $BVRI$ and $g'r'i'z'$ filter sets between 2015 May and 2015 December from 10 different member observatories: Canela's Robotic Observatory (CROW), Kutztown University Observatory (KUO), the University of Louisville's Moore Observatory Ritchey-Chr\'{e}tien (MORC) telescope, the Manner-Vanderbilt Ritchey-Chr\'{e}tien (MVRC) telescope located on the Mt. Lemmon summit of Steward Observatory, Brigham Young University's Orson Pratt Observatory (Pratt), Westminster College Observatory (WCO), Wellesley College's Whitin Observatory (Whitin), and Brigham Young University's West Mountain Observatory (WMO).
The transit observations are summarized in Table \ref{tab:Photom} and the light curves shown in Figures \ref{fig:LCs1a} and \ref{fig:LCs1b}. 
Figure \ref{fig:LCs2} shows all follow-up light curves combined and binned in 5 minute intervals. This combined and binned light curve is not used for analysis, but rather to showcase the overall statistical power of the follow-up photometry. 

All observatories were Network Time Protocol (NTP) or GPS synchronized to sub-millisecond accuracy, and all times have been converted to barycentric Julian dates at mid-exposure, \bjdtdb \citep{Eastman:2010}. All data were processed using the AstroImageJ package (AIJ; \citealt{Collins:2016, Collins:2013}).\footnote{\url{http://www.astro.louisville.edu/software/astroimagej}}

\begin{table*}
 \footnotesize
 \centering
 \setlength\tabcolsep{2pt}
 \caption{Photometric follow-up observations of the KELT-16\MakeLowercase{b} transit}
 \begin{tabular}{l		c			r@{.}l				c			r@{}l				c		c		c			r@{}l				r			r			c		c		c
}
\hline																																										
\hline																																										
Observatory 	 & 	Location	 & 		\multicolumn{2}{c}{Aperture}			 & 	 Date 	 & 		\multicolumn{2}{c}{Transit}			 & 	 Filter 	 & 	 FOV 	 & 	 Pixel Scale 	 & 		\multicolumn{2}{c}{FWHM$^b$} 			 & 	 \multicolumn{1}{c}{Exposure}		 & 	\multicolumn{1}{c}{Cycle$^c$}		 & 	Duty Cycle$^d$	 & 	rms$^e$	 & 	PNR$^f$	 \\	    
    	 & 		 & 		\multicolumn{2}{c}{(m)}			 & 	 (UT 2015) 	 & 		\multicolumn{2}{c}{Epoch$^a$}			 & 	 	 & 	 ($\arcmin \times \arcmin$) 	 & 	 ($\arcsec$pixel$^\textrm{-1}$) 	 & 		\multicolumn{2}{c}{($\arcsec$)}			 & 	 \multicolumn{1}{c}{(s)} 		 & 	 \multicolumn{1}{c}{(s)}		 & 	(\%)	 & 	(10$^\textrm{-3}$)	 & 	(10$^\textrm{-3}\,$min$^\textrm{-1}$)	 \\	    
\hline																																										
    MVRC 	 & 	Ariz.	 & 	\hspace{6pt}	0	 & 	6	 & 	 May 23 	 & 	\hspace{3pt}	-	 & 	84	 & 	 $r'$ 	 & 	 $26 \times 26$ 	 & 	0.39	 & 	\hspace{6pt}	5	 & 		 & 	100	\hspace{8pt}	 & 	112	\hspace{2pt}	 & 	89	 & 	1.4	 & 	2.0	 \\	    
    KUO 	 & 	Pa.	 & 	\hspace{6pt}	0	 & 	6	 & 	 May 24 	 & 	\hspace{3pt}	-	 & 	83	 & 	 $V$ 	 & 	 $19.5 \times 13$ \hspace{3pt} 	 & 	0.76	 & 	\hspace{6pt}	4	 & 	.2	 & 	120	\hspace{8pt}	 & 	143	\hspace{2pt}	 & 	84	 & 	1.8	 & 	2.7	 \\ 	    
    KUO 	 & 	Pa.	 & 	\hspace{6pt}	0	 & 	6	 & 	 May 24 	 & 	\hspace{3pt}	-	 & 	83	 & 	 $I$ 	 & 	 $19.5 \times 13$ \hspace{3pt} 	 & 	0.76	 & 	\hspace{6pt}	4	 & 	.2	 & 	120	\hspace{8pt}	 & 	143	\hspace{2pt}	 & 	84	 & 	1.9	 & 	3.0	 \\ 	    
    KUO 	 & 	Pa.	 & 	\hspace{6pt}	0	 & 	6	 & 	 May 26 	 & 	\hspace{3pt}	-	 & 	81	 & 	 $B$ 	 & 	 $19.5 \times 13$ \hspace{3pt} 	 & 	0.76	 & 	\hspace{6pt}	4	 & 	.4	 & 	165	\hspace{8pt}	 & 	184	\hspace{2pt}	 & 	90	 & 	3.1	 & 	5.4	 \\ 	    
    Pratt 	 & 	Utah	 & 	\hspace{6pt}	0	 & 	4	 & 	 Jun 23 	 & 	\hspace{3pt}	-	 & 	52	 & 	 $R$ 	 & 	 $25 \times 25$ 	 & 	0.37	 & 	\hspace{6pt}	3	 & 	.6	 & 	90	\hspace{8pt}	 & 	109	\hspace{2pt}	 & 	82	 & 	3.7	 & 	5.0	 \\	    
    WMO 	 & 	Utah	 & 	\hspace{6pt}	0	 & 	9	 & 	 Jun 25 	 & 	\hspace{3pt}	-	 & 	50	 & 	 $B$ 	 & 	 $20.5 \times 20.5$ 	 & 	0.61	 & 	\hspace{6pt}	2	 & 	.6	 & 	40	\hspace{8pt}	 & 	55	\hspace{2pt}	 & 	73	 & 	2.2	 & 	2.1	 \\	    
    WMO 	 & 	Utah	 & 	\hspace{6pt}	0	 & 	9	 & 	 Jun 25 	 & 	\hspace{3pt}	-	 & 	50	 & 	 $V$ 	 & 	 $20.5 \times 20.5$ 	 & 	0.61	 & 	\hspace{6pt}	2	 & 	.6	 & 	30	\hspace{8pt}	 & 	45	\hspace{2pt}	 & 	67	 & 	2.0	 & 	1.8	 \\	    
    WMO 	 & 	Utah	 & 	\hspace{6pt}	0	 & 	9	 & 	 Jun 25 	 & 	\hspace{3pt}	-	 & 	50	 & 	 $R$ 	 & 	 $20.5 \times 20.5$ 	 & 	0.61	 & 	\hspace{6pt}	2	 & 	.6	 & 	20	\hspace{8pt}	 & 	35	\hspace{2pt}	 & 	57	 & 	2.4	 & 	1.8	 \\	    
    WMO 	 & 	Utah	 & 	\hspace{6pt}	0	 & 	9	 & 	 Jun 25 	 & 	\hspace{3pt}	-	 & 	50	 & 	 $I$ 	 & 	 $20.5 \times 20.5$ 	 & 	0.61	 & 	\hspace{6pt}	2	 & 	.6	 & 	20	\hspace{8pt}	 & 	43	\hspace{2pt}	 & 	47	 & 	3.0	 & 	2.6	 \\	    
    MORC 	 & 	Ky.	 & 	\hspace{6pt}	0	 & 	6	 & 	 Jul 25 	 & 	\hspace{3pt}	-	 & 	19	 & 	 $z'$ 	 & 	 $26 \times 26$ 	 & 	0.39	 & 	\hspace{6pt}	5	 & 		 & 	240	\hspace{8pt}	 & 	259	\hspace{2pt}	 & 	93	 & 	1.9	 & 	4.0	 \\ 	    
    CROW 	 & 	Portugal	 & 	\hspace{6pt}	0	 & 	25	 & 	 Aug 4 	 & 	\hspace{3pt}	-	 & 	\hspace{2pt} 9	 & 	 $R_{\textrm{\tiny{C}}}$ 	 & 	 $28 \times 19$ 	 & 	1.11	 & 	\hspace{6pt}	5	 & 	.3	 & 	200	\hspace{8pt}	 & 	252	\hspace{2pt}	 & 	79	 & 	1.8	 & 	3.7	 \\ 	    
    Whitin 	 & 	Mass.	 & 	\hspace{6pt}	0	 & 	6	 & 	 Nov 2 	 & 	\hspace{3pt}		 & 	84	 & 	 $i'$ 	 & 	 $20 \times 20$ 	 & 	0.58	 & 	\hspace{6pt}	4	 & 	.4	 & 	80	\hspace{8pt}	 & 	109	\hspace{2pt}	 & 	73	 & 	2.5	 & 	3.4	 \\	    
    WCO 	 & 	Pa.	 & 	\hspace{6pt}	0	 & 	35	 & 	 Nov 3 	 & 	\hspace{3pt}		 & 	85	 & 	 $r'$ 	 & 	 $24 \times 16$ 	 & 	1.44	 & 	\hspace{6pt}	3	 & 	.8	 & 	160	\hspace{8pt}	 & 	182	\hspace{2pt}	 & 	88	 & 	1.8	 & 	3.1	 \\	    
    MORC 	 & 	Ky.	 & 	\hspace{6pt}	0	 & 	6	 & 	 Nov 4 	 & 	\hspace{3pt}		 & 	86	 & 	 $g'$ 	 & 	 $26 \times 26$ 	 & 	0.39	 & 	\hspace{6pt}	9	 & 		 & 	100	\hspace{8pt}	 & 	125	\hspace{2pt}	 & 	80	 & 	1.4	 & 	2.1	 \\	    
    MORC 	 & 	Ky.	 & 	\hspace{6pt}	0	 & 	6	 & 	 Nov 4 	 & 	\hspace{3pt}		 & 	86	 & 	 $i'$ 	 & 	 $26 \times 26$ 	 & 	0.39	 & 	\hspace{6pt}	9	 & 		 & 	100	\hspace{8pt}	 & 	125	\hspace{2pt}	 & 	80	 & 	1.2	 & 	1.7	 \\	    
    KUO 	 & 	Pa.	 & 	\hspace{6pt}	0	 & 	6	 & 	 Nov 4 	 & 	\hspace{3pt}		 & 	86	 & 	 $V$ 	 & 	 $19.5 \times 13$ \hspace{3pt} 	 & 	0.76	 & 	\hspace{6pt}	3	 & 	.8	 & 	120	\hspace{8pt}	 & 	144	\hspace{2pt}	 & 	83	 & 	1.5	 & 	2.4	 \\ 	    
    KUO 	 & 	Pa.	 & 	\hspace{6pt}	0	 & 	6	 & 	 Nov 4 	 & 	\hspace{3pt}		 & 	86	 & 	 $I$ 	 & 	 $19.5 \times 13$ \hspace{3pt} 	 & 	0.76	 & 	\hspace{6pt}	3	 & 	.8	 & 	120	\hspace{8pt}	 & 	144	\hspace{2pt}	 & 	83	 & 	1.8	 & 	2.8	 \\ 	    
    Whitin 	 & 	Mass.	 & 	\hspace{6pt}	0	 & 	6	 & 	 Dec 7 	 & 	\hspace{3pt}	1	 & 	20	 & 	 $r'$ 	 & 	 $20 \times 20$ 	 & 	0.58	 & 	\hspace{6pt}	4	 & 	.0	 & 	80	\hspace{8pt}	 & 	93	\hspace{2pt}	 & 	86	 & 	3.0	 & 	3.7	 \\	    
    Whitin 	 & 	Mass.	 & 	\hspace{6pt}	0	 & 	6	 & 	 Dec 7 	 & 	\hspace{3pt}	1	 & 	20	 & 	 $i'$ 	 & 	 $20 \times 20$ 	 & 	0.58	 & 	\hspace{6pt}	4	 & 	.0	 & 	80	\hspace{8pt}	 & 	93	\hspace{2pt}	 & 	86	 & 	3.1	 & 	3.9	 \\	    
\hline																																										
\hline																																										

\end{tabular}
\begin{flushleft}
  \footnotesize{ \textbf{\textsc{NOTES:}} $^a$The zeroth epoch is set to be the mid-transit time, $T_\textrm{C}$, of the global fit to the radial velocity (RV) data (see Table \ref{tab:KELT-16b_1}). $^b$The average full width at half maximum (FWHM) of the stellar point spread function (PSF). $^c$The cycle time is the mean of exposure time plus dead time during periods of back-to-back exposures. $^d$The duty cycle is the fraction of cycle time spent exposing. $^e$The rms is the root mean square scatter of the residuals to the best fit transit model. $^f$The photometric noise rate (PNR) is calculated as rms$/\sqrt{\Gamma}$, where $\Gamma$ is the mean number of cycles per minute (adapted from \citealt{Fulton:2011}).}
\end{flushleft}
\label{tab:Photom}
\end{table*}

\subsection{Spectroscopic Follow-up}
\label{sec:Spectra}

We used the Tillinghast Reflector Echelle Spectrograph (TRES; \citealt{Szentgyorgyi:2007, Furesz:2008}), on the 1.5 m telescope at the Fred Lawrence Whipple Observatory (FLWO) on Mt. Hopkins, Arizona, to obtain high resolution spectra. We obtained a total of 20 spectra between UT 2015 May 29 and 2015 December 8. The spectra have an average signal-to-noise ratio per resolution element (SNRe) of 39 at the peak of the continuum near the Mg b triplet at 519 nm. The spectra have a resolving power of $R \sim 44,000$ and were extracted as described by \citet{Buchave:2010}.

We derive relative radial velocities (RVs) by cross-correlating each observed spectrum order by order against the strongest observed spectrum from the wavelength range 4460 - 6280 \AA. The observation used as the template has an RV of 0 \kms \hspace{2 pt} by definition. The data are reported in Table \ref{tab:Spectra}.

The absolute RV of the system is determined using the absolute velocity of the star in the spectrum with the highest SNRe, which is also the spectrum used as the template when deriving the relative RVs. The absolute RV is adjusted to the International Astronomical Union (IAU) Radial Velocity Standard Star system \citep{Stefanik:1999} by first taking the RV determined by cross-correlating the spectrum with the best synthetic spectrum match near the Mg b order (see \S\ref{sec:SpecParams} for more details) and then adding the gamma velocity from the orbital solution using the relative RVs. A correction of -0.61 \kms  -- determined from extensive observations of the IAU standard stars -- is included to correct for the non-inclusion of the gravitational redshift in the synthetic spectrum. The absolute velocity of the star is found to be 
-29.7 \kms \hspace{2pt}  $\pm$ 0.1 \kms, where the uncertainty is an estimate of the residual systematic errors in the IAU system.

\begin{figure}
\includegraphics[width=1\linewidth]{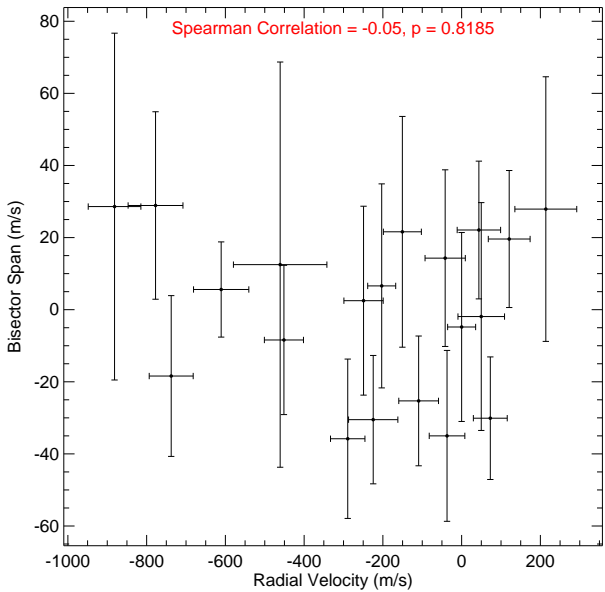}
\vspace{-0.2in}
\caption{\footnotesize Bisector spans for the TRES RV spectra for KELT-16 plotted against the RV values.  There is no correlation between these quantities.}
\label{fig:Bisectors}
\end{figure}

We derive values for the line profile bisector spans using procedures outlined in \citet{Buchave:2010}. The bisector values are reported in Table \ref{tab:Spectra} and shown in Figure \ref{fig:Bisectors}. We measure line bisector spans to check for variations from background blends \citep{Mandushev:2005} or star spots \citep{Queloz:2001}. We expect bisector variations caused from these astrophysical phenomena to vary in phase with the RV variations. There is no indication that the periodic signal is due to any astrophysical phenomena other than the orbital motion.

\begin{table}
\footnotesize
 \centering
 \setlength\tabcolsep{3pt}
\caption{TRES Relative Radial Velocity Measurements}
\begin{tabular}{c r@{.}l c r@{.}l c}
 \hline
 \hline
  \bjdtdb &  \multicolumn{2}{c}{\hspace{4pt} RV} & RV error & \multicolumn{2}{c}{Bisector} & Bisector Error \\
          & \multicolumn{2}{c}{\hspace{4pt} (\ms)} & (\ms) & \multicolumn{2}{c}{(\ms)} & (\ms) \\
 \hline
2457171.92841 & \hspace{6pt} -777&4 & 50.5 & \hspace{2pt} 28&9 & 26.0 \\
2457203.85614 & \hspace{6pt} -881&5 & 48.5 & \hspace{2pt} 28&6 & 48.1 \\
2457323.60011 & \hspace{6pt}  213&8 & 57.0 & \hspace{2pt} 27&9 & 36.7 \\
2457345.65687 & \hspace{6pt} -224&6 & 45.4 & \hspace{2pt} -30&5 & 17.8 \\
2457346.62517 & \hspace{6pt} -249&1 & 36.2 & \hspace{2pt} 2&5 & 26.2 \\
2457347.67298 & \hspace{6pt}  -37&1 & 32.8 & \hspace{2pt} -35&0 & 23.7 \\
2457348.65228 & \hspace{6pt}    0&0 & 25.8 & \hspace{2pt} -4&8 & 26.2 \\
2457349.63889 & \hspace{6pt}   43&8 & 40.0 & \hspace{2pt} 22&1 & 19.1 \\
2457350.60616 & \hspace{6pt}   72&8 & 31.2 & \hspace{2pt} -30&1 & 17.0 \\
2457351.58680 & \hspace{6pt}  120&8 & 38.5 & \hspace{2pt} 19&6 & 19.0 \\
2457354.63919 & \hspace{6pt}   49&8 & 42.9 & \hspace{2pt} -1&9 & 31.6 \\
2457355.64983 & \hspace{6pt}  -41&8 & 37.1 & \hspace{2pt} 14&3 & 24.5 \\
2457356.62390 & \hspace{6pt} -109&2 & 36.6 & \hspace{2pt} -25&3 & 18.0 \\
2457357.65057 & \hspace{6pt} -203&0 & 25.8 & \hspace{2pt} 6&6 & 28.3 \\
2457358.60143 & \hspace{6pt} -150&5 & 35.0 & \hspace{2pt} 21&6 & 32.0 \\
2457360.59367 & \hspace{6pt} -289&3 & 31.7 & \hspace{2pt} -35&8 & 22.1 \\
2457361.59428 & \hspace{6pt} -451&3 & 36.0 & \hspace{2pt} -8&4 & 20.7 \\
2457362.64672 & \hspace{6pt} -610&7 & 50.9 & \hspace{2pt} 5&6 & 13.2 \\
2457363.63593 & \hspace{6pt} -460&9 & 86.1 & \hspace{2pt} 12&5 & 56.2 \\
2457364.61912 & \hspace{6pt} -737&6 & 40.5 & \hspace{2pt} -18&4 & 22.3 \\
\hline
 \hline
\end{tabular}
 \label{tab:Spectra}
\begin{flushleft}
\end{flushleft}
\end{table}

\begin{figure}
\includegraphics[width=1\linewidth]{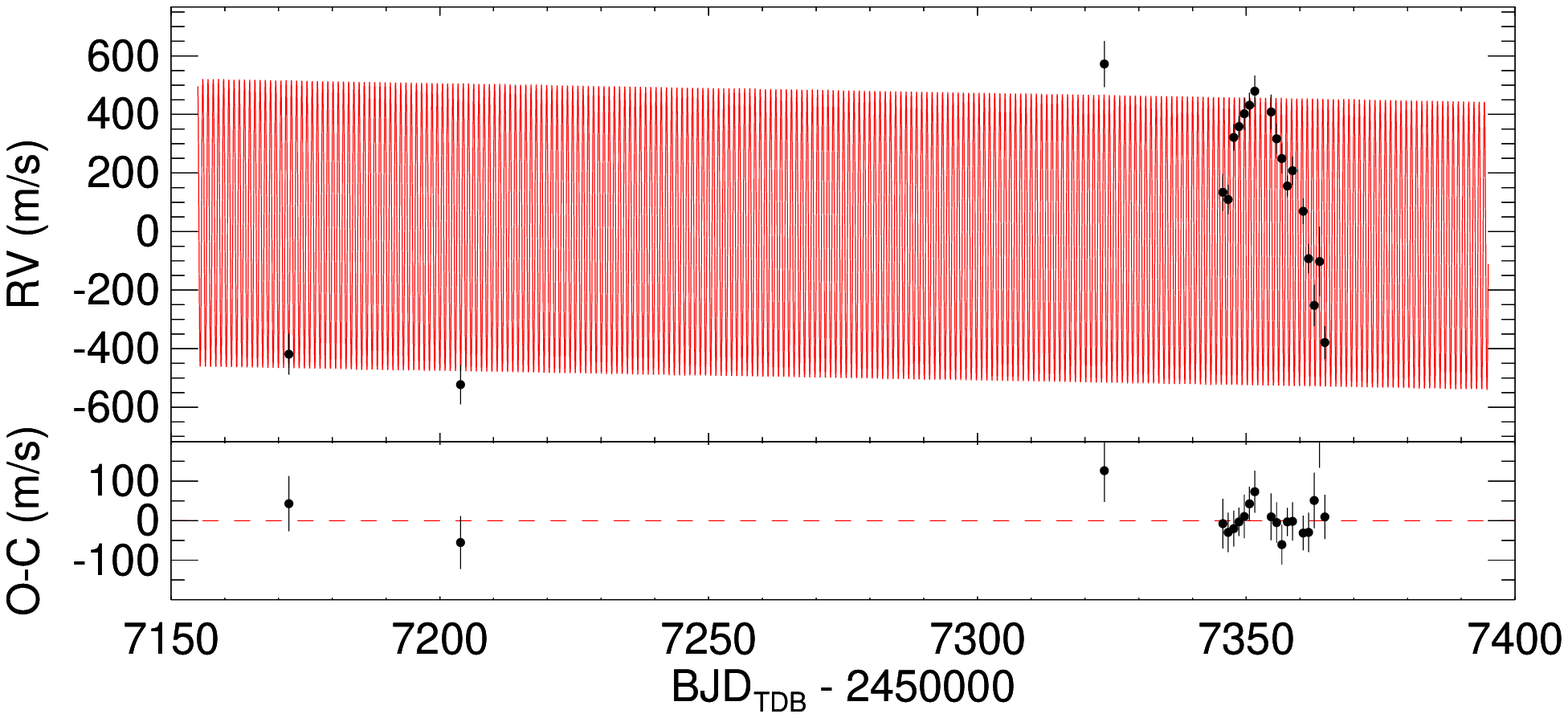}
   \vspace{-.3in}
\includegraphics[width=1\linewidth]{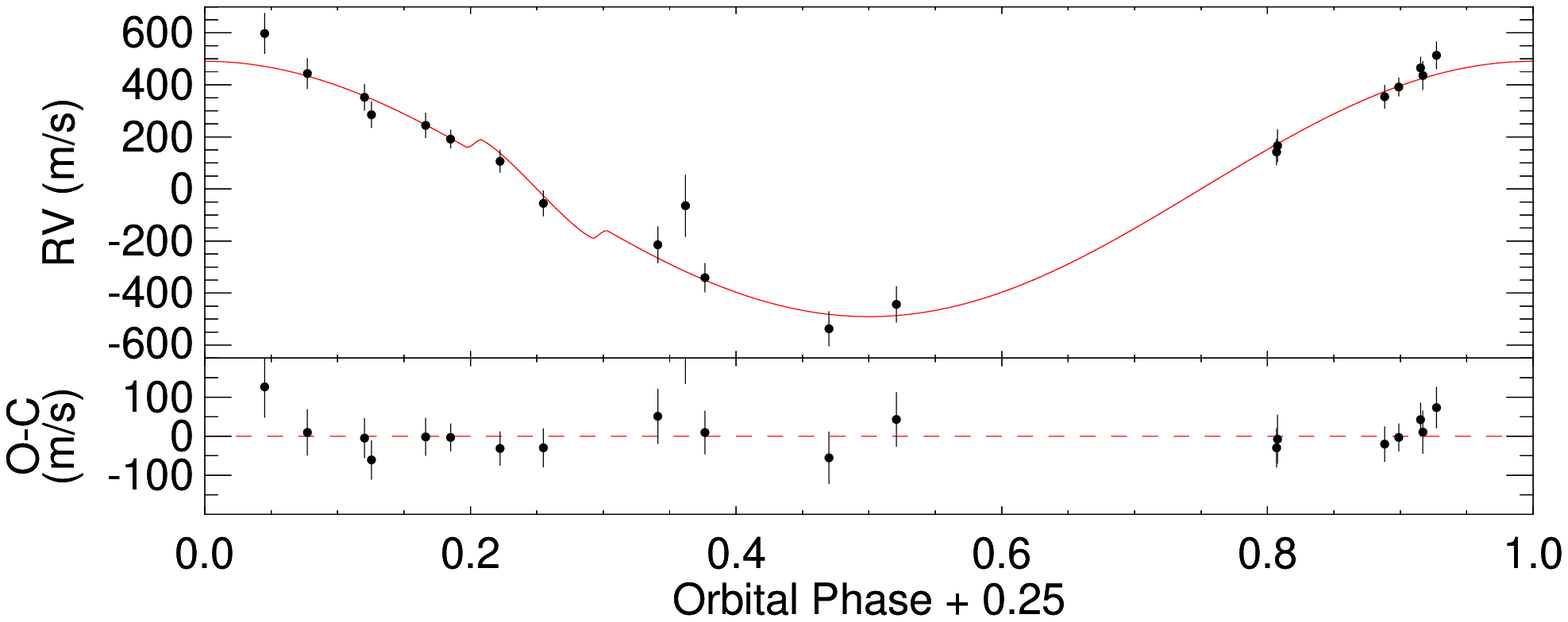}
   \vspace{.1in}
\caption{\footnotesize \textit{Top}: Relative radial velocity (RV) measurements of KELT-16 from TRES (black data points and error bars). The red curve represents the global model fit to the data (\S\ref{sec:GlobalFit}). \textit{Bottom}: The TRES RV measurements phased to the global fit-determined period of 0.9689951 days. The O-C residuals are shown in the panels directly below each plot. Note that one point (at phase $+ 0.25$ of $\sim 0.36$) falls slightly outside the range of displayed residuals for the bottom plot.}
\label{fig:RVs} 
\end{figure}

%
%
%


\subsection{High-Contrast Imaging}
\label{sec:AO}

\begin{figure}
\includegraphics[width=1\linewidth]{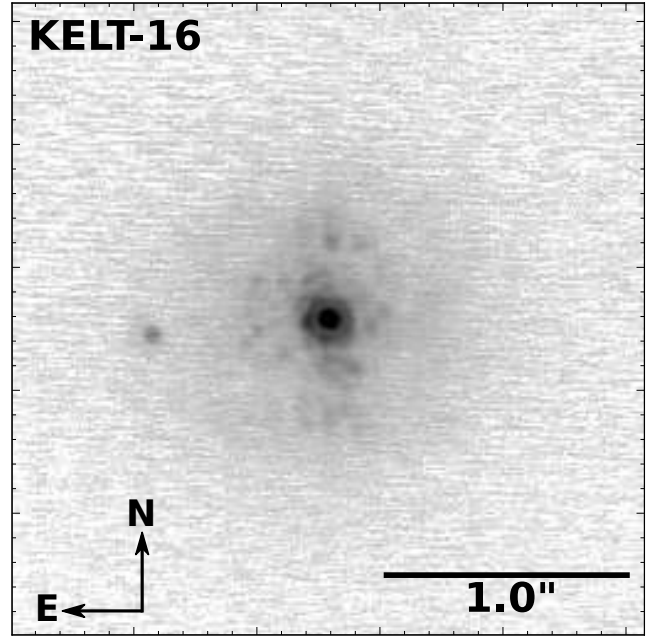}
\caption{\footnotesize NIRC2/Keck II AO $K_\textrm{S}$ stacked image of KELT-16 on a logarithmic scale. A stellar companion can be seen $\sim 0.7\arcsec$ to the east of the primary star. This image is for display purposes and includes only the 3 best frames; all 12 frames are used in the analysis.}
\label{fig:AO}
\end{figure}

We obtained natural guide star adaptive optics (AO) images of KELT-16 on 2015 December 27 with the NIRC2 instrument \citep{Matthews:1994} on Keck II. Twelve 15 s exposures were taken in the $K_\textrm{S}$ band using the full NIRC2 array (1024 $\times$ 1024 pixels) with the narrow camera setting ($0\farcs010$ pixel$^{-1}$). The full width at half maximum (FWHM) of the target star PSF was $\sim$ 5 pixels = $0\farcs050$. We calibrate and remove image artifacts using dome flat fields and dark frames. Figure \ref{fig:AO} shows a stacked image of the system using the three best frames, while all 12 frames are used for the analysis. A stellar companion is detected roughly $0\farcs7$ east of the target star.

We measure the flux ratio and on-sky separation of the companion by fitting a two-peak PSF to the data. Each peak is modeled as a combination of a Moffat and Gaussian function within an aperture of radius 2 $\times$ FWHM. Details on the PSF fitting routine can be found in \citet{Ngo:2015}. We integrate the best-fit PSF over the same aperture to calculate the flux ratio and calculate the difference between centroids to find the separation and position angle (PA) of the companion. For the position measurement, we apply the new NIRC2 astrometric corrections from \citet{Service:2016} to correct for the NIRC2 array's distortion and rotation. We find the flux ratio of the primary to secondary to be $56.5 \pm 5.5$, or $\Delta K_\textrm{S} = 4.4 \pm 0.1$ magnitudes. Adding this to the 2MASS-measured $K_\textrm{S}$ of the primary (Table \ref{tab:LitProps}) yields a secondary brightness of $K_\textrm{S} = 15.0 \pm 0.1$ magnitudes. The companion is separated from the KELT-16 primary by $0.7177\arcsec \pm 0.0015\arcsec$ at a PA of $95\fdg16 \pm 0\fdg22$.

\begin{figure}
\includegraphics[width=1\linewidth]{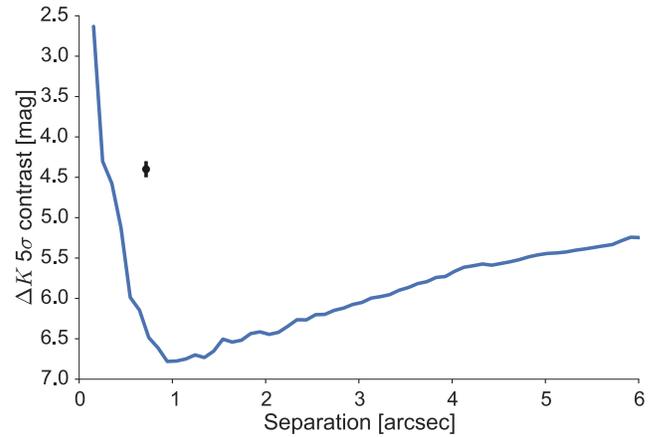}
\caption{\footnotesize NIRC2/Keck II AO $5\sigma$ contrast curve, showing the limiting magnitude for a companion detection in our observations. The data point and error bars mark the observation of the KELT-16 companion (the horizontal error bars are smaller than the data point itself). The sensitivity decreases for separations $\gtrsim 1\arcsec$ due to the decreasing overlap of frames farther from the primary in the dithering pattern employed: the 12 observed frames overlap fully within $\sim 1\arcsec$ of the primary star, while the outer regions are covered by only either four or eight frames. This strategy provides the highest sensitivity in the innermost $\sim 1\arcsec$, which is usually the most interesting region, while maintaining the ability to detect somewhat brighter companions at larger separations.}
\label{fig:Contrast}
\end{figure}

We also compute a $5\sigma$ contrast curve. To do so, we divide the stacked image into a series of annuli of width equal to the primary star's FWHM, where each annulus is used to compute our sensitivity at a given distance from the primary star. Then, for every pixel, we compute the sum of the flux of all neighboring pixels within a FWHM $\times$ FWHM = 5 pixel $\times$ 5 pixel box. The standard deviation of these values within the same annulus is the $1\sigma$ contrast for that annulus. We divide the limiting contrast flux by the total flux in a 5 pixel $\times$ 5 pixel box centered on the primary star to get a magnitude difference. Figure \ref{fig:Contrast} shows the resulting $5\sigma$ contrast curve in $\Delta K_\textrm{S}$.   

%
%
%

\section{Analysis and Results}
\label{sec:Analysis}

\subsection{Stellar Parameters from Spectra}
\label{sec:SpecParams}

We obtain stellar parameters from the observed TRES spectra (\S\ref{sec:Spectra}) using the Stellar Parameter Classification (SPC) tool \citep{Buchave:2012}. SPC cross-correlates an observed spectrum against a grid of synthetic spectra based on Kurucz atmospheric models \citep{Kurucz:1992, Kurucz:1979}. The weighted average results are: $T_\textrm{eff} = 6227 \pm 55$ K, log($g_\star$) $= 4.03 \pm 0.11$, $\meh = -0.01 \pm 0.08$, and $v\sin{i_\star} = 7.6 \pm 0.5$ \kms. These values were calculated by taking an average of the stellar parameters determined for each spectrum individually and then weighting them according to the cross-correlation function peak height.

\subsection{EXOFAST Global Fit}
\label{sec:GlobalFit}

To determine the system parameters for KELT-16 and place it in context with other known exoplanet systems, we conduct a global fit of our follow-up photometric and spectroscopic observations. The global fit uses a modified version of EXOFAST which we refer to as multi-EXOFAST \citep{Eastman:2013}\footnote{Multi-EXOFAST has not yet been updated with the new conversion constants for solar and planetary properties from IAU 2015 Resolution B3 \citep{Mamajek:2015}, but the differences are within the uncertainties of our results.} to constrain $R_\star$ and $M_\star$ using either the Yonsei-Yale (YY) stellar evolution models \citep{Demarque:2004} or the empirical Torres relations \citep{Torres:2010}. The RV measurements from TRES (\S\ref{sec:Spectra}) and raw KELT-FUN follow-up light curves (with detrending parameters -- see \citealt{Collins:2014}) are used as inputs for the fit, and the \feh \hspace{2pt} and $T_\textrm{eff}$  (with errors) determined from the SPC analysis of the TRES spectra (\S\ref{sec:SpecParams}) and the period determined from the follow-up light curves are used as priors. The blended flux from KELT-16b's nearby companion as determined by spectral energy distribution (SED) modeling (\S\ref{sec:SED}) is subtracted from each follow-up light curve. (See \citealt{Siverd:2012} for a more detailed description of the global fit procedure.)

For both the YY models and the Torres relations, we run a global fit with eccentricity constrained to zero and another with eccentricity as a free parameter, for a total of four fits. The results are shown in Tables \ref{tab:KELT-16b_1} and \ref{tab:KELT-16b_2}. All four fits are consistent with each other to 1$\sigma$; for our interpretation and discussion in this paper, we adopt the YY circular fit. The best-fitting linear ephemeris is $T_\textrm{C} = 2457247.24791\pm0.00019$ \bjdtdb and $P = 0.9689951 \pm 0.0000024$ d. Using the standards of \citet{Pecaut:2013}, the best-fitting $T_\textrm{eff}$ corresponds to a stellar spectral type of F7V.  

\begin{table*}
 \scriptsize
\centering
\setlength\tabcolsep{1.5pt}
\caption{Median values and 68\% confidence intervals for the physical and orbital parameters of the KELT-16 system}
  \label{tab:KELT-16b_1}
  \begin{tabular}{lccccc}
  \hline
  \hline
   Parameter & Units & \textbf{Adopted Value} & Value & Value & Value \\
   & & \textbf{(YY circular)} & (YY eccentric) & (Torres circular) & (Torres eccentric)\\
 \hline
\multicolumn{2}{l}{Stellar Parameters} & & & &\\
                               ~~~$M_{\star}$\dotfill &Mass (\msun)\dotfill & $1.211_{-0.046}^{+0.043}$&$1.206_{-0.048}^{+0.047}$&$1.232_{-0.058}^{+0.060}$&$1.224_{-0.059}^{+0.061}$\\
                             ~~~$R_{\star}$\dotfill &Radius (\rsun)\dotfill & $1.360_{-0.053}^{+0.064}$&$1.329_{-0.070}^{+0.073}$&$1.390_{-0.062}^{+0.063}$&$1.355_{-0.076}^{+0.077}$\\
                         ~~~$L_{\star}$\dotfill &Luminosity (\lsun)\dotfill & $2.52_{-0.22}^{+0.27}$&$2.40_{-0.27}^{+0.29}$&$2.62_{-0.25}^{+0.27}$&$2.49_{-0.29}^{+0.31}$\\
                             ~~~$\rho_{\star}$\dotfill &Density (cgs)\dotfill & $0.679_{-0.081}^{+0.079}$&$0.72_{-0.10}^{+0.12}$&$0.647_{-0.070}^{+0.084}$&$0.694_{-0.094}^{+0.12}$\\
                  ~~~$\log{g_\star}$\dotfill &Surface gravity (cgs)\dotfill & $4.253_{-0.036}^{+0.031}$& $4.272_{-0.042}^{+0.043}$&$4.242_{-0.032}^{+0.034}$&$4.262_{-0.041}^{+0.043}$\\
                  ~~~$\teff$\dotfill &Effective temperature (K)\dotfill & $6236\pm54$&$6237_{-55}^{+54}$&$6235\pm53$&$6235_{-54}^{+53}$\\
                                 ~~~$\feh$\dotfill &Metallicity\dotfill & $-0.002_{-0.085}^{+0.086}$&$0.002_{-0.085}^{+0.086}$&$0.002_{-0.079}^{+0.078}$&$0.002_{-0.079}^{+0.080}$\\
\hline
 \multicolumn{2}{l}{Planet Parameters} & & & & \\
                                   ~~~$e$\dotfill &Eccentricity\dotfill & --- &$0.034_{-0.025}^{+0.047}$& --- &$0.031_{-0.022}^{+0.045}$\\
        ~~~$\omega_\star$\dotfill &Argument of periastron (degrees)\dotfill & --- &$-65_{-55}^{+43}$& --- &$-64_{-67}^{+52}$\\
                                  ~~~$P$\dotfill &Period (days)\dotfill & $0.9689951\pm0.0000024$&$0.9689951_{-0.0000024}^{+0.0000025}$&$0.9689951\pm0.0000024$&$0.9689951\pm0.0000025$\\
                           ~~~$a$\dotfill &Semi-major axis (AU)\dotfill & $0.02044_{-0.00026}^{+0.00024}$&$0.02041_{-0.00028}^{+0.00026}$&$0.02055\pm0.00033$&$0.02051\pm0.00033$\\
                                 ~~~$M_\textrm{P}$\dotfill &Mass (\mj)\dotfill & $2.75_{-0.15}^{+0.16}$&$2.78_{-0.17}^{+0.18}$&$2.79_{-0.16}^{+0.17}$&$2.80_{-0.18}^{+0.19}$\\
                               ~~~$R_\textrm{P}$\dotfill &Radius (\rj)\dotfill & $1.415_{-0.067}^{+0.084}$&$1.384_{-0.080}^{+0.089}$&$1.452_{-0.079}^{+0.081}$&$1.415_{-0.089}^{+0.093}$\\
                           ~~~$\rho_\textrm{P}$\dotfill &Density (cgs)\dotfill & $1.20\pm0.18$&$1.30_{-0.23}^{+0.27}$&$1.452_{-0.079}^{+0.081}$&$1.22_{-0.21}^{+0.27}$\\
                      ~~~$\log{g_\textrm{P}}$\dotfill &Surface gravity\dotfill & $3.530_{-0.049}^{+0.042}$&$3.554\pm0.058$&$3.515\pm0.046$&$3.539_{-0.057}^{+0.059}$\\
               ~~~$T_\textrm{eq}$\dotfill &Equilibrium temperature (K)\dotfill & $2453_{-47}^{+55}$&$2427\pm64$&$2472\pm53$&$2443_{-66}^{+64}$\\
                           ~~~$\Theta$\dotfill &Safronov number\dotfill & $0.0654\pm0.0045$&$0.0677_{-0.0057}^{+0.0063}$&$0.0639_{-0.0044}^{+0.0046}$&$0.0662_{-0.0056}^{+0.0063}$\\
                   ~~~$\fave$\dotfill &Incident flux (\fluxcgs)\dotfill & $8.22_{-0.61}^{+0.77}$&$7.86_{-0.81}^{+0.87}$&$8.48_{-0.70}^{+0.75}$&$8.07_{-0.85}^{+0.89}$\\
 \hline
 \multicolumn{2}{l}{RV Parameters} & & & & \\
       ~~~$T_\textrm{C}$\dotfill &Time of inferior conjunction (\bjdtdb)\dotfill & $2457247.24791\pm0.00019$&$2457247.24791\pm0.00019$&$2457247.24791\pm0.00019$&$2457247.24791\pm0.00019$\\
               ~~~$T_\textrm{P}$\dotfill &Time of periastron (\bjdtdb)\dotfill & --- &$2457246.84_{-0.15}^{+0.12}$& --- &$2457246.84_{-0.18}^{+0.15}$\\
                        ~~~$K$\dotfill &RV semi-amplitude (m/s)\dotfill & $494\pm25$&$501\pm29$&$494\pm25$&$500\pm29$\\
                    ~~~$M_\textrm{P}\sin{i}$\dotfill &Minimum mass (\mj)\dotfill & $2.74\pm0.15$&$2.76_{-0.17}^{+0.18}$&$2.77_{-0.16}^{+0.17}$&$2.79\pm0.18$\\
                           ~~~$M_\textrm{P}/M_{\star}$\dotfill &Mass ratio\dotfill & $0.00217\pm0.00011$&$0.00220\pm0.00013$&$0.00216\pm0.00011$&$0.00219\pm0.00013$\\
                       ~~~$u$\dotfill &RM linear limb darkening\dotfill & $0.6051_{-0.0059}^{+0.0063}$&$0.6052_{-0.0059}^{+0.0064}$&$0.6052_{-0.0058}^{+0.0063}$&$0.6054_{-0.0059}^{+0.0063}$\\
                                ~~~$\gamma_\textrm{~TRES}$\dotfill &m/s\dotfill & $-353\pm31$&$-359_{-35}^{+34}$&$-353\pm31$&$-358\pm34$\\
                  ~~~$\dot{\gamma}$\dotfill &RV slope (m/s/day)\dotfill & $-0.37\pm0.34$&$-0.37\pm0.37$&$-0.37\pm0.34$&$-0.37\pm0.37$\\
                                                           ~~~$\ecosw$\dotfill & \dotfill & --- &$0.012_{-0.017}^{+0.026}$& --- &$0.010_{-0.016}^{+0.026}$\\
                                                        ~~~$\esinw$\dotfill & \dotfill & --- &$-0.024_{-0.048}^{+0.026}$& --- &$-0.019_{-0.048}^{+0.022}$\\
 \hline
 \hline
 \end{tabular}
\end{table*}

\begin{table*}
\scriptsize
 \centering
\setlength\tabcolsep{1.5pt}
\caption{Median values and 68\% confidence intervals for the physical and orbital parameters of the KELT-16 system (continued)}
  \label{tab:KELT-16b_2}
  \begin{tabular}{lccccc}
  \hline
  \hline
   Parameter & Units & \textbf{Adopted Value} & Value & Value & Value \\
   & & \textbf{(YY circular)} & (YY eccentric) & (Torres circular) & (Torres eccentric)\\
 \hline
 \hline
 \multicolumn{2}{l}{Primary Transit} & & & & \\
~~~$R_\textrm{P}/R_{\star}$\dotfill &Radius of the planet in stellar radii\dotfill & $0.1070_{-0.0012}^{+0.0013}$&$0.1072_{-0.0012}^{+0.0013}$&$0.1074\pm0.0013$&$0.1074\pm0.0013$\\
           ~~~$a/R_\star$\dotfill &Semi-major axis in stellar radii\dotfill & $3.23_{-0.13}^{+0.12}$&$3.30_{-0.16}^{+0.17}$&$3.18_{-0.12}^{+0.13}$&$3.26_{-0.15}^{+0.17}$\\
                          ~~~$i$\dotfill &Inclination (degrees)\dotfill & $84.4_{-2.3}^{+3.0}$&$84.4_{-2.0}^{+2.7}$&$83.5_{-1.9}^{+2.7}$&$83.8_{-1.9}^{+2.6}$\\
                               ~~~$b$\dotfill &Impact parameter\dotfill & $0.32_{-0.16}^{+0.11}$&$0.332_{-0.15}^{+0.098}$&$0.359_{-0.14}^{+0.087}$&$0.360_{-0.14}^{+0.087}$\\
                             ~~~$\delta$\dotfill &Transit depth\dotfill & $0.01146_{-0.00025}^{+0.00029}$&$0.01148_{-0.00026}^{+0.00028}$&$0.01154_{-0.00027}^{+0.00028}$&$0.01154_{-0.00027}^{+0.00028}$\\
                    ~~~$T_\textrm{FWHM}$\dotfill &FWHM duration (days)\dotfill & $0.09237\pm0.00044$&$0.09244\pm0.00044$&$0.09248\pm0.00044$&$0.09248\pm0.00044$\\
              ~~~$\tau$\dotfill &Ingress/egress duration (days)\dotfill & $0.01133_{-0.00099}^{+0.0013}$&$0.0115_{-0.0010}^{+0.0012}$&$0.0118_{-0.0011}^{+0.0012}$&$0.0118_{-0.0011}^{+0.0012}$\\
                     ~~~$T_{14}$\dotfill &Total duration (days)\dotfill & $0.1037_{-0.0010}^{+0.0013}$&$0.1039_{-0.0011}^{+0.0012}$&$0.1043\pm0.0012$&$0.1043\pm0.0012$\\
   ~~~$P_\textrm{~T}$\dotfill &A priori non-grazing transit probability\dotfill & $0.2764_{-0.0097}^{+0.012}$&$0.265_{-0.023}^{+0.018}$&$0.281_{-0.011}^{+0.010}$&$0.269_{-0.023}^{+0.018}$\\
             ~~~$P_\textrm{~T,G}$\dotfill &A priori transit probability\dotfill & $0.343_{-0.013}^{+0.015}$&$0.328_{-0.028}^{+0.023}$&$0.348\pm0.014$&$0.334_{-0.028}^{+0.023}$\\
               ~~~$T_\textrm{~C,0}$\dotfill &Mid-transit time (\bjdtdb)\dotfill & $2457165.85179\pm0.00049$&$2457165.85182\pm0.00049$&$2457165.85179\pm0.00049$&$2457165.85181\pm0.00049$\\
               ~~~$T_\textrm{~C,1}$\dotfill &Mid-transit time (\bjdtdb)\dotfill & $2457166.82114_{-0.0010}^{+0.00097}$&$2457166.8212\pm0.0010$&$2457166.82115_{-0.0010}^{+0.00100}$&$2457166.82119_{-0.0010}^{+0.00100}$\\
               ~~~$T_\textrm{~C,2}$\dotfill &Mid-transit time (\bjdtdb)\dotfill & $2457166.8240_{-0.0014}^{+0.0016}$&$2457166.8240_{-0.0013}^{+0.0016}$&$2457166.8240_{-0.0013}^{+0.0015}$&$2457166.8241_{-0.0013}^{+0.0015}$\\
               ~~~$T_\textrm{~C,3}$\dotfill &Mid-transit time (\bjdtdb)\dotfill & $2457168.7572_{-0.0019}^{+0.0016}$&$2457168.7572_{-0.0019}^{+0.0016}$&$2457168.7571_{-0.0019}^{+0.0016}$&$2457168.7571_{-0.0019}^{+0.0016}$\\
               ~~~$T_\textrm{~C,4}$\dotfill &Mid-transit time (\bjdtdb)\dotfill & $2457196.8608_{-0.0012}^{+0.0011}$&$2457196.8608_{-0.0012}^{+0.0011}$&$2457196.8608_{-0.0012}^{+0.0011}$&$2457196.8608_{-0.0012}^{+0.0011}$\\
               ~~~$T_\textrm{~C,5}$\dotfill &Mid-transit time (\bjdtdb)\dotfill & $2457198.7984\pm0.0011$&$2457198.7984\pm0.0011$& $2457198.7984\pm0.0011$&$2457198.7984\pm0.0011$\\
               ~~~$T_\textrm{~C,6}$\dotfill &Mid-transit time (\bjdtdb)\dotfill & $2457198.79803_{-0.00066}^{+0.00065}$&$2457198.79806_{-0.00067}^{+0.00066}$&$2457198.79804\pm0.00066$&$2457198.79807_{-0.00067}^{+0.00066}$\\
               ~~~$T_\textrm{~C,7}$\dotfill &Mid-transit time (\bjdtdb)\dotfill & $2457198.79914_{-0.00087}^{+0.00085}$&$2457198.79918_{-0.00086}^{+0.00084}$&$2457198.79914_{-0.00085}^{+0.00083}$&$2457198.79917_{-0.00085}^{+0.00083}$\\
               ~~~$T_\textrm{~C,8}$\dotfill &Mid-transit time (\bjdtdb)\dotfill & $2457198.7969_{-0.0014}^{+0.0015}$&$2457198.7969\pm0.0013$&$2457198.7969_{-0.0012}^{+0.0013}$&$2457198.7969_{-0.0012}^{+0.0013}$\\
               ~~~$T_\textrm{~C,9}$\dotfill &Mid-transit time (\bjdtdb)\dotfill & $2457228.8368_{-0.0011}^{+0.0010}$&$2457228.8368_{-0.0011}^{+0.0010}$&$2457228.83676_{-0.0010}^{+0.00099}$&$2457228.83676_{-0.0010}^{+0.00100}$\\
              ~~~$T_\textrm{~C,10}$\dotfill &Mid-transit time (\bjdtdb)\dotfill & $2457238.52801_{-0.00075}^{+0.00077}$&$2457238.52804_{-0.00074}^{+0.00077}$&$2457238.52803_{-0.00074}^{+0.00076}$&$2457238.52807_{-0.00075}^{+0.00076}$\\
              ~~~$T_\textrm{~C,11}$\dotfill &Mid-transit time (\bjdtdb)\dotfill & $2457328.64376_{-0.00078}^{+0.00080}$&$2457328.64380_{-0.00078}^{+0.00081}$&$2457328.64377_{-0.00079}^{+0.00082}$&$2457328.64381_{-0.00080}^{+0.00082}$\\
              ~~~$T_\textrm{~C,12}$\dotfill &Mid-transit time (\bjdtdb)\dotfill & $2457329.61152_{-0.00066}^{+0.00067}$&$2457329.61156\pm0.00067$&$2457329.61155_{-0.00066}^{+0.00068}$&$2457329.61158_{-0.00067}^{+0.00068}$\\
              ~~~$T_\textrm{~C,13}$\dotfill &Mid-transit time (\bjdtdb)\dotfill & $2457330.58170_{-0.00061}^{+0.00060}$&$2457330.58173\pm0.00061$&$2457330.58172_{-0.00061}^{+0.00060}$&$2457330.58174_{-0.00062}^{+0.00061}$\\
              ~~~$T_\textrm{~C,14}$\dotfill &Mid-transit time (\bjdtdb)\dotfill & $2457330.58146\pm0.00056$&$2457330.58150_{-0.00057}^{+0.00056}$&$2457330.58147\pm0.00056$&$2457330.58150\pm0.00056$\\
              ~~~$T_\textrm{~C,15}$\dotfill &Mid-transit time (\bjdtdb)\dotfill & $2457330.58154_{-0.00069}^{+0.00067}$&$2457330.58157_{-0.00070}^{+0.00068}$&$2457330.58153_{-0.00069}^{+0.00067}$&$2457330.58157_{-0.00070}^{+0.00069}$\\
              ~~~$T_\textrm{~C,16}$\dotfill &Mid-transit time (\bjdtdb)\dotfill & $2457330.5826_{-0.0010}^{+0.0011}$&$2457330.5826_{-0.0010}^{+0.0011}$&$2457330.5827\pm0.0010$&$2457330.5827\pm0.0010$\\
              ~~~$T_\textrm{~C,17}$\dotfill &Mid-transit time (\bjdtdb)\dotfill & $2457363.5265\pm0.0011$&$2457363.5266\pm0.0011$&$2457363.5265\pm0.0011$&$2457363.5266\pm0.0011$\\
              ~~~$T_\textrm{~C,18}$\dotfill &Mid-transit time (\bjdtdb)\dotfill & $2457363.5278\pm0.0014$&$2457363.5278\pm0.0014$&$2457363.5278_{-0.0014}^{+0.0013}$&$2457363.5279_{-0.0014}^{+0.0013}$\\
                     ~~~$u_\textrm{1~B}$\dotfill &Linear Limb-darkening\dotfill & $0.544_{-0.012}^{+0.013}$&$0.543_{-0.012}^{+0.014}$&$0.544_{-0.012}^{+0.014}$&$0.544_{-0.012}^{+0.013}$\\
                  ~~~$u_\textrm{2~B}$\dotfill &Quadratic Limb-darkening\dotfill & $0.2356_{-0.0090}^{+0.0078}$&$0.2359_{-0.0091}^{+0.0078}$&$0.2354_{-0.0090}^{+0.0078}$&$0.2355_{-0.0090}^{+0.0079}$\\
                     ~~~$u_\textrm{1~I}$\dotfill &Linear Limb-darkening\dotfill & $0.2231_{-0.0055}^{+0.0064}$&$0.2235_{-0.0056}^{+0.0065}$&$0.2231_{-0.0056}^{+0.0065}$&$0.2235_{-0.0056}^{+0.0065}$\\
                  ~~~$u_\textrm{2~I}$\dotfill &Quadratic Limb-darkening\dotfill & $0.3045_{-0.0031}^{+0.0028}$&$0.3044_{-0.0031}^{+0.0028}$&$0.3047_{-0.0029}^{+0.0026}$&$0.3045_{-0.0029}^{+0.0026}$\\
                     ~~~$u_\textrm{1~R}$\dotfill &Linear Limb-darkening\dotfill & $0.2939_{-0.0064}^{+0.0076}$&$0.2941_{-0.0065}^{+0.0077}$&$0.2939_{-0.0064}^{+0.0077}$&$0.2942_{-0.0065}^{+0.0077}$\\
                  ~~~$u_\textrm{2~R}$\dotfill &Quadratic Limb-darkening\dotfill & $0.3148_{-0.0030}^{+0.0026}$&$0.3148_{-0.0030}^{+0.0026}$&$0.3149_{-0.0029}^{+0.0025}$&$0.3148_{-0.0029}^{+0.0025}$\\
                ~~~$u_\textrm{1~Sloan~g'}$\dotfill &Linear Limb-darkening\dotfill & $0.4743_{-0.0100}^{+0.011}$&$0.4742_{-0.0100}^{+0.012}$&$0.4746_{-0.0099}^{+0.011}$&$0.475_{-0.010}^{+0.011}$\\
             ~~~$u_\textrm{2~Sloan~g'}$\dotfill &Quadratic Limb-darkening\dotfill & $0.2706_{-0.0064}^{+0.0051}$&$0.2708_{-0.0065}^{+0.0051}$&$0.2704_{-0.0064}^{+0.0051}$&$0.2705_{-0.0064}^{+0.0052}$\\
                ~~~$u_\textrm{1~Sloan~i'}$\dotfill &Linear Limb-darkening\dotfill & $0.2412_{-0.0057}^{+0.0066}$&$0.2415_{-0.0057}^{+0.0067}$&$0.2411_{-0.0057}^{+0.0067}$&$0.2416_{-0.0058}^{+0.0067}$\\
             ~~~$u_\textrm{2~Sloan~i'}$\dotfill &Quadratic Limb-darkening\dotfill & $0.3064\pm0.0028$&$0.3063_{-0.0029}^{+0.0028}$&$0.3066_{-0.0027}^{+0.0026}$&$0.3064_{-0.0028}^{+0.0026}$\\
                ~~~$u_\textrm{1~Sloan~r'}$\dotfill &Linear Limb-darkening\dotfill & $0.3145_{-0.0067}^{+0.0079}$&$0.3147_{-0.0068}^{+0.0080}$&$0.3146_{-0.0067}^{+0.0080}$&$0.3148_{-0.0068}^{+0.0080}$\\
             ~~~$u_\textrm{2~Sloan~r'}$\dotfill &Quadratic Limb-darkening\dotfill & $0.3161_{-0.0031}^{+0.0025}$&$0.3161_{-0.0031}^{+0.0024}$&$0.3162_{-0.0030}^{+0.0024}$&$0.3161_{-0.0030}^{+0.0024}$\\
                ~~~$u_\textrm{1~Sloan~z'}$\dotfill &Linear Limb-darkening\dotfill & $0.1899_{-0.0051}^{+0.0059}$&$0.1901_{-0.0052}^{+0.0060}$&$0.1898_{-0.0051}^{+0.0059}$&$0.1902_{-0.0052}^{+0.0059}$\\
             ~~~$u_\textrm{2~Sloan~z'}$\dotfill &Quadratic Limb-darkening\dotfill & $0.2979_{-0.0034}^{+0.0026}$&$0.2978_{-0.0034}^{+0.0026}$&$0.2981_{-0.0031}^{+0.0025}$&$0.2979_{-0.0032}^{+0.0025}$\\
                     ~~~$u_\textrm{1~V}$\dotfill &Linear Limb-darkening\dotfill & $0.3787_{-0.0076}^{+0.0090}$&$0.3788_{-0.0077}^{+0.0091}$&$0.3788_{-0.0076}^{+0.0090}$&$0.3790_{-0.0077}^{+0.0091}$\\
                  ~~~$u_\textrm{2~V}$\dotfill &Quadratic Limb-darkening\dotfill & $0.3033_{-0.0039}^{+0.0027}$&$0.3034_{-0.0039}^{+0.0027}$&$0.3033_{-0.0039}^{+0.0027}$&$0.3033_{-0.0039}^{+0.0027}$\\
                  \hline
\multicolumn{2}{l}{Secondary Eclipse} & & & & \\
                  ~~~$T_\textrm{S}$\dotfill &Time of eclipse (\bjdtdb)\dotfill & $2457246.76341\pm0.00019$&$2457246.771_{-0.010}^{+0.016}$&$2457246.76341\pm0.00019$&$2457246.7694_{-0.0098}^{+0.016}$\\
                  ~~~$b_\textrm{~S}$\dotfill &Impact parameter\dotfill & --- &$0.309_{-0.14}^{+0.096}$& --- &$0.339_{-0.13}^{+0.088}$\\
                  ~~~$T_\textrm{S,FWHM}$\dotfill &FWHM duration (days)\dotfill & --- &$0.0887_{-0.0072}^{+0.0041}$& --- &$0.0896_{-0.0071}^{+0.0035}$\\
            ~~~$\tau_\textrm{~S}$\dotfill &Ingress/egress duration (days)\dotfill & --- &$0.0107_{-0.0012}^{+0.0014}$& --- &$0.0111_{-0.0013}^{+0.0014}$\\
                   ~~~$T_\textrm{S,14}$\dotfill &Total duration (days)\dotfill & --- &$0.0996_{-0.0084}^{+0.0051}$& --- &$0.1009_{-0.0083}^{+0.0044}$\\
   ~~~$P_\textrm{~S}$\dotfill &A priori non-grazing eclipse probability\dotfill & --- &$0.279_{-0.010}^{+0.011}$& --- &$0.282\pm0.011$\\
             ~~~$P_\textrm{~S,G}$\dotfill &A priori eclipse probability\dotfill & --- &$0.346_{-0.014}^{+0.015}$& --- &$0.350_{-0.015}^{+0.014}$\\
     \hline
 \hline
\end{tabular}
 \begin{flushleft}
  \footnotesize \textbf{\textsc{NOTES:}} The mid-transit times $T_\textrm{C,0}$ through $T_\textrm{C,18}$ correspond to those of the 19 KELT-FUN light curves in the same order as they are listed in Table \ref{tab:Photom}.
  \end{flushleft}
\end{table*}

\subsection{Evolutionary Analysis}
\label{sec:HRD}

\begin{figure}
\includegraphics[width=0.75\columnwidth,angle=90]{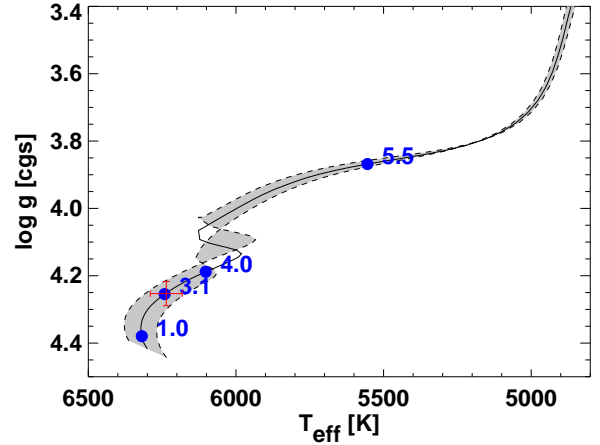}
\caption{\footnotesize
KELT-16 Yonsei-Yale (YY) model fit on an HR diagram. Plotted are KELT-16's $T_\textrm{eff}$ and log($g_\star$) (red point and error bars), the YY evolutionary track best fit by KELT-16's $M_\star$ and [Fe/H] (solid black line), and select points of isochrone intersection with the evolutionary track in units of Gyr (blue points). $T_\textrm{eff}$, log($g_\star$), [Fe/H], and $M_\star$ are all taken from the YY circular orbit case of the global fit (Table \ref{tab:KELT-16b_1}). The tracks for the extreme range of 1$\sigma$ uncertainties on $M_\star$ and [Fe/H] are shown as dashed lines bracketing the grey swath (which becomes disjoined near the ``blue hook'' due to the overlap of evolutionary tracks there). The best fit age is found to be 3.1 $\pm$ 0.3 Gyr. 
}
\label{fig:HRD}
\end{figure}

We estimate the age of KELT-16 by fitting the YY stellar evolution model to the stellar $T_\textrm{eff}$, log($g_\star$), \feh, and $M_\star$ determined from the circular YY case of our global fit (\S\ref{sec:GlobalFit} and Table \ref{tab:KELT-16b_1}). The fitting process can be understood graphically as finding the intersection point on an HR diagram of the evolutionary track that best fits $M$ and \feh with the associated isochrone that best fits $T_\textrm{eff}$ and log($g$). The result is an age of 3.1 $\pm$ 0.3 Gyr, indicating that KELT-16 is undergoing core hydrogen fusion and is slightly past the midpoint of its lifespan on the main sequence (Figure \ref{fig:HRD}). The uncertainty in age reflects only the propagation of the uncertainties in $T_\textrm{eff}$, log($g_\star$), \feh, and $M_\star$ from the global fit, and does not include systematic or calibration uncertainties of the YY model itself. 


\subsection{SED Analysis}
\label{sec:SED}

We estimate the distance and reddening to KELT-16 by fitting the \citet{Kurucz:1979, Kurucz:1992} stellar atmosphere models to the SED. We use broad-band photometry data from the literature spanning the GALEX $NUV$ band at 0.227 $\mu$m to the $WISE3$ band at 11.6 $\mu$m (Table \ref{tab:LitProps}). 

However, the detection of a stellar companion by AO imaging (\S\ref{sec:AO}) indicates that the photometry in all of these passbands is blended. The AO imaging was single-banded, measuring only the $K_\textrm{S}$-band flux ratio of the two stars. But under the assumption that the two stars are bound, we can iteratively use the distance and reddening from a preliminary SED fit (uncorrected for blending) along with the best-fit isochronal age (\S\ref{sec:HRD}) to determine a $T_\textrm{eff}$ for the secondary star based on the model of \citet{Baraffe:2015}. This $T_\textrm{eff}$ is then used to extrapolate the $K_\textrm{S}$-band flux ratio to the other photometric bands in order to separate the flux contributions from the two stars (see \S\ref{sec:Companion}).

\begin{figure}
  \centering \includegraphics[width=0.75\columnwidth, angle=90]{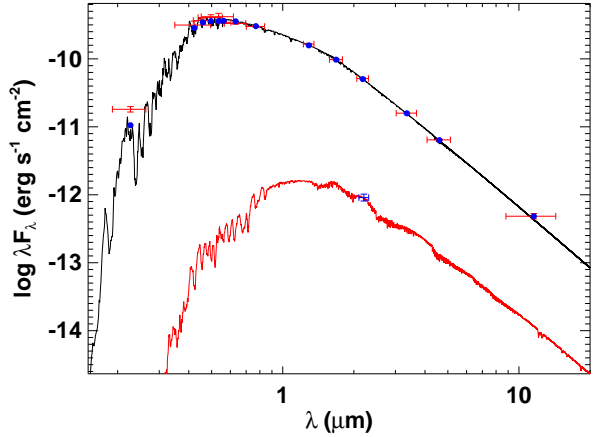}
  \caption{\footnotesize \textit{Upper curve}: SED of KELT-16. The red data points are the measured fluxes of KELT-16 in the passbands listed in Table \ref{tab:LitProps} after correcting for blending from the companion star. The vertical error bars are the 1$\sigma$ photometric uncertainties whereas the horizontal error bars are the effective widths of the passbands. The black solid curve is the best-fit theoretical SED from the models of \citet{Kurucz:1979, Kurucz:1992} with $T_\textrm{eff}$ fixed at 6227 K. The blue dots are the predicted passband-integrated fluxes of the SED fit in the observed photometric bands. \textit{Lower curve}: SED of KELT-16's stellar companion. The blue data point and error bars represent the $K_\textrm{S}$ band flux of the companion as measured by AO imaging. 
  The red solid curve is the best-fit theoretical SED from the models of \citet{Kurucz:1979, Kurucz:1992} with $T_\textrm{eff}$ fixed at 3416 K, where this temperature relies on the assumption that the companion is bound to KELT-16.}
  \label{fig:SED}
\end{figure}

Figure \ref{fig:SED} shows the final best fit stellar atmospheres for both the KELT-16 primary and secondary stars. For these fits, we fixed the $T_\textrm{eff}$ of the primary as determined directly from spectroscopic observations (\S\ref{sec:SpecParams}) and the $T_\textrm{eff}$ of the secondary as determined from the Baraffe model and leave distance, $d$, and reddening, $A_V$, as free parameters. We find best fit values of $d = 399 \pm 19$ pc and $A_V = 0.04 \pm 0.04$ magnitudes, where the uncertainties reflect only the propagation of the uncertainties of the measured fluxes and do not include systematic or calibration uncertainties of the stellar atmosphere models themselves. 

\subsection{Stellar Companion}
\label{sec:Companion}

A stellar companion was detected near KELT-16 by high-contrast AO imaging (\S\ref{sec:AO}). First, we compute the likelihood that the two stars are gravitationally bound rather than a chance alignment of foreground and background objects. 

\begin{figure}
\centering
\includegraphics[width=1\columnwidth,angle=0]{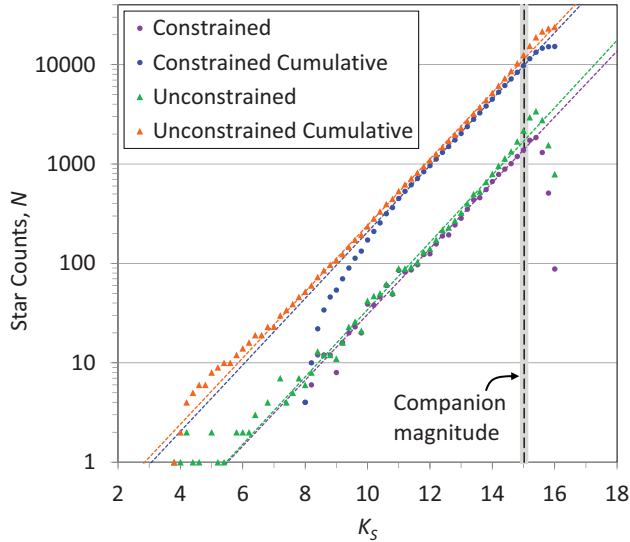}
\caption{\footnotesize Absolute and cumulative stellar number density observed by 2MASS as a function of $K_\textrm{S}$ magnitude in a 1 deg$^2$ circular region centered on KELT-16. Star counts are grouped in bins of 0.2 magnitudes corresponding to the $\pm 1 \sigma$ uncertainty of the companion star's brightness. We include both the ``unconstrained'' case in which all 2MASS catalog entries are counted and the case in which counts are constrained by excluding those entries having quality flags (see text). The black vertical dashed line and grey swath mark the $K_\textrm{S} = 15.0 \pm 0.1$ brightness of the companion. The colored dashed lines are model fits to the data of the form $y=ae^{bx}$, excluding the brighter data points due to low number statistics and excluding the fainter tails due to incompleteness.}
\label{fig:StarCounts}
\end{figure}

Figure \ref{fig:StarCounts} shows the absolute and cumulative stellar number density observed by the 2 Micron All-Sky Survey (2MASS; \citealt{Skrutskie:2006}) as a function of $K_\textrm{S}$ magnitude in a 1 deg$^2$ circular region surrounding KELT-16. We include both the ``unconstrained'' case in which all 2MASS catalog entries are counted and the case in which counts are constrained by quality flags.\footnote{2MASS entries are excluded if they have quality flags for asteroid or comet association (Aflg = 0), extended source contamination (Xflg = 0), artifact or confusion contamination (Cflg = *0), poor photometric quality (Rflg = *2), or poor photometric fit (Bflg = *1).} Regardless of constraints, the $K_\textrm{S} = 15.0 \pm 0.1$ companion is found to be safely brighter than the 2MASS faintness limit of $K_\textrm{S} \sim 15.5$ for this region of sky. Within this limit, 2MASS star counts have a very high level of completeness, and can thus be used directly to calculate a  model-independent alignment probability.

In particular, we calculate the probability that the AO observations would detect \textit{by chance} a star which is at least as near to the primary as the detected companion and at least as bright as the detected companion (restricting also by the measured PA of the detected companion is not necessary since the AO imaging covered all PAs). 
To do so, we determine the fractional area of the sky occupied by $0\farcs72$ radii circles surrounding all 2MASS sources with brightnesses of $K_\textrm{S} \leq 15.0$. 
Technically, this $K_\textrm{S}$ bound should be taken as a function of separation to ensure it remains above the $5\sigma$ contrast curve (Figure \ref{fig:Contrast}), but neglecting this only makes our calculation more conservative, since it will cause the probability of chance alignment at lesser separations to be overestimated. This approach assumes that the 2MASS catalog is complete to the faintness limit, that catalog entries have at least a $1\farcs44$ separation, and that there is no blending in the 2MASS reported fluxes. Although none of these are strictly true, they are very good approximations: 2MASS completeness in the $K_\textrm{S}$-band is reported as 99.56 \% over the entire sky and likely to be higher in the region immediately surrounding KELT-16 due to its lack of bright targets or tiling gaps and modest angular separation from the Galactic plane ($b \sim -9\degr$); 2MASS spatial resolution was seeing-limited and is reported to average $\sim$ 4"; and the quality flags for confusion and photometric fit help to discount blended targets \citep{Skrutskie:2006}.

Even in the most conservative case, unconstrained by (i.e. ignoring) quality flags, we count 12,487 2MASS sources of $K_\textrm{S} \leq 15.0$. The probability of a chance alignment of one of these stars within $0\farcs72$ of the target is 0.1569 \%, resulting in a 99.8431 \% or $\sim 3.2 \sigma$ confidence that the two stars are bound.

If the two stars are bound, we can compute a number of model-dependent stellar and orbital properties of the secondary. We start with the directly observed $K_\textrm{S}$ flux ratio and angular separation and assume the same isochrone-modeled age and SED-modeled distance and extinction as the primary star (\S\ref{sec:HRD} and \S\ref{sec:SED}). From these values we calculate the extinction-corrected absolute $K_\textrm{S}$ magnitude, which, when coupled with the age, can be fit to the models of \citet{Baraffe:2015} to obtain $T_\textrm{eff}$, $L$, $R$, $M$, and log($g$). This $T_\textrm{eff}$ can then be matched to a spectral type and colors -- and thus apparent magnitudes across multiple bands -- using the spectral typing standards of \citet{Pecaut:2013}. In short, we find the companion to be a $V = 19.6$ M3V-type red dwarf. The full stellar properties of the secondary are summarized in Table \ref{tab:Companion}.

\begin{table}
 \scriptsize{
 \centering
 \caption{Properties of the KELT-16 Stellar Companion}
 \label{tab:Companion}
 \begin{tabular}{llcc}
    \hline
    \hline
Parameter & Description & Value & Source \\
\hline
\multicolumn{4}{l}{Directly Observed Parameters}                                            \\
$K_\textrm{\tiny{S,1}}/K_\textrm{\tiny{S,2}}$\dotfill & Flux ratio\dotfill & $56.5 \pm 5.5$    & AO obs.                  \\
    $\Delta r$\dotfill    & Angular separation (mas)\dotfill        & $717.7 \pm 1.5$ & AO obs.           \\
PA  \dotfill           & Position angle (degrees)\dotfill           & $95.16 \pm 0.22$  & AO obs.           \\
\hline
\multicolumn{4}{l}{Modeled Stellar Parameters}                                              \\
$B$\dotfill            & Johnson magnitude (mag)\dotfill            & $21.2 \pm 0.1$  & PM13              \\
$V$\dotfill            & Johnson magnitude (mag)\dotfill            & $19.6 \pm 0.1$  & PM13              \\
$J$\dotfill            & 2MASS magnitude (mag) \dotfill             & $15.9 \pm 0.1$  & PM13              \\
$H$\dotfill            & 2MASS magnitude (mag) \dotfill             & $15.3 \pm 0.1$  & PM13              \\
$K_\textrm{S}$\dotfill          & 2MASS magnitude (mag) \dotfill             & $15.0 \pm 0.1$  & AO \&             \\
               &                                    &                   & \hspace{3pt} 2MASS obs. \\
Age\dotfill            & Age (Gyr)\dotfill                          & $3.1 \pm 0.3$     & Isochrone fit     \\
$d$\dotfill            & Distance (pc)\dotfill                      & $399 \pm 19$      & SED fit           \\
$A_V$\dotfill          & Extinction coeff. (mag)\dotfill            & $0.04 \pm 0.04$   & SED fit           \\
$M_{K_\textrm{S}}$\dotfill    & Abs. 2MASS mag. (mag)\dotfill              & $7.0 \pm 0.2$   & AO obs. \&        \\
               &                                    &                   & \hspace{3pt} SED fit \\
$T_\textrm{\tiny{eff}}$\dotfill          & Effective temperature (K)\dotfill          & $3420 \pm 70$     & BHAC15            \\
$L_\star$\dotfill      & Luminosity ($L_\odot$)\dotfill                   & $0.0107	\pm 0.0037$ & BHAC15        \\
$R_\star$\dotfill      & Radius ($R_\odot$)\dotfill                       & $0.295 \pm 0.040$ & BHAC15            \\
$M_\star$\dotfill      & Mass ($M_\odot$)\dotfill                         & $0.300 \pm 0.050$ & BHAC15            \\
log($g$)\dotfill       & Surface gravity (cgs)\dotfill              & $4.97 \pm 0.05$   & BHAC15            \\
\feh\dotfill           & Metallicity\dotfill                      & $-0.002 \pm 0.086$ & Global fit        \\
Spectral Type\dotfill  & Spectral type\dotfill                      & M3V               & PM13              \\
\hline
\multicolumn{4}{l}{Modeled Orbital Parameters}                                              \\
$a_{\textrm{\tiny{min}},e=0}$\dotfill  & Minimum circular\hspace{15pt}\dotfill    & $286 \pm 14$      & AO obs. \&        \\
                              & \hspace{3pt} semi-major axis (AU) & & \hspace{3pt} SED fit  \\
$P_{\textrm{\tiny{min}},e=0}$\dotfill & Minimum circular\dotfill    & $3940 \pm 280$    & Global fit \&     \\
                              & \hspace{3pt} period (yr) &              & \hspace{3pt} BHAC15  \\
$a_{\textrm{\tiny{min}},e\rightarrow1}$\dotfill & Minimum eccentric\hspace{11pt}\dotfill & $143 \pm 7$ & AO obs. \&      \\
                                        & \hspace{3pt} semi-major axis (AU) & & \hspace{3pt} SED fit \\ 
$P_{\textrm{\tiny{min}},e\rightarrow1}$\dotfill & Minimum eccentric\dotfill  & $1390 \pm 100$ & Global fit \& \\
                                        & \hspace{3pt} period (yr) &    & \hspace{3pt} BHAC15 \\
    \hline
    \hline
 \end{tabular}
 \begin{flushleft}
  \footnotesize \textbf{\textsc{NOTES:}} All values except those directly observed by NIRC2/Keck II AO rely on the assumption that the companion is bound. All values that depend on the Baraffe et al. (2015; BHAC15) model also depend on the direct AO observations and both the isochrone and SED model fits. All values that depend on Pecaut \& Mamajek (2013; PM13) standards also depend on the $T_\textrm{eff}$ from BHAC15. The 68 \% confidence intervals account only for the propagation of measurement uncertainties and do not include any systematic or calibration uncertainty inherent in the models themselves.
  \end{flushleft}
  }
\end{table} 

The stars' directly observed angular separation and SED-modeled distance imply a current minimum separation of $286 \pm 14$ AU, or higher if the primary and secondary are not in the same sky plane. From this we calculate the minimum orbital period via Kepler's third law using the primary mass as determined by the global fit (\S\ref{sec:GlobalFit}) and secondary mass as determined by the \citet{Baraffe:2015} model. We find that the minimum orbital period could range from $3940 \pm 280$ yr for a circular orbit to $1390 \pm 100$ yr for a maximum eccentricity orbit ($\lim_{e \to 1}$). (Note that the distribution of observed stellar binary eccentricities has been found to be independent of stellar properties and is nearly flat, although it is also possible to fit it with a broad Gaussian of $e_\textrm{avg}$ $\sim$ 0.4; \citealt{Duchene:2013}.)

\begin{figure}
\centering
\includegraphics[width=1\columnwidth,angle=0]{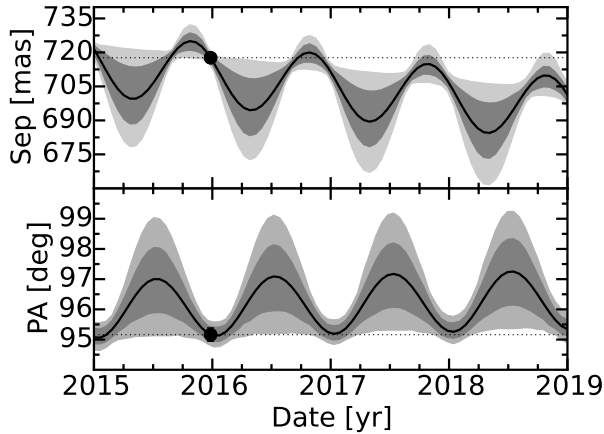}
\caption{\footnotesize
Predicted separation in milliarcseconds (top panel) and position angle, PA, in degrees (bottom panel) of the KELT-16 stellar primary and secondary. The two data points show the NIRC2/Keck II AO observations on 2015 Dec 27 presented in this work. The dotted lines -- which show no change in separation or PA -- are for the case in which the stars are bound and do not account for their mutual orbital motion, which is negligible on the few year timescale of this plot relative to their estimated $> 1000$ yr minimum orbital period. The solid curves are for the case in which the two stars are unbound and assumes the secondary is at infinity (i.e. has no motion), with the sinusoidal variations being due to the parallax of the primary and the linear trends being due to the proper motion of the primary as measured by NOMAD (Table \ref{tab:LitProps}). The sinusoidal amplitude is slightly overestimated since the secondary is at a finite distance. The grey swaths estimate the 1 and 2 $\sigma$ uncertainties of NIRC2/Keck II AO observations.
}
\label{fig:ProperMotion}
\end{figure}

It is statistically likely that the stars are bound, but this argument could be strengthened with high contrast imaging in multiple bands (e.g. $J-K_\textrm{S}$) to confirm whether or not the two stars have similar photometric distances, and/or high contrast imaging in a future epoch to confirm whether or not the two stars have common parallax and proper motion. A search of major catalogs turns up no observations of the companion in past epochs (not surprising given its faintness and small separation), and additional observations were not able to be obtained by the time of publication (but are being actively pursued). Figure \ref{fig:ProperMotion} shows a model of the two stars' separation and PA with NIRC2/Keck II AO observation uncertainties, and reveals that the soonest opportunity to check for a separation difference at the $\sim 3 \sigma$ level will be in mid-2018. The PA difference will not exceed $\sim 2 \sigma$ in the near future.

\subsection{UVW Space Motion}
\label{sec:UVW}

We examine the three-dimensional space motion of KELT-16 to determine whether its kinematics match those of one of the main stellar populations of the Galaxy. The absolute RV measured from the TRES spectra of KELT-16 is $-29.7 \pm 0.1$ \kms \hspace{2pt} (\S\ref{sec:Spectra}) and the proper motions from NOMAD are $\mu_{\alpha} = 5.1 \pm 0.7$ mas yr$^{-1}$ and $\mu_{\delta} = 0.4 \pm 0.7$ mas yr$^{-1}$ (Table \ref{tab:LitProps}). Adopting the SED-modeled distance (\S\ref{sec:SED}) and local standard of rest (LSR) as defined in \citet{Coskunoglu:2011}, these values transform to $U$, $V$, and $W$ space motions of $-5.4 \pm 1.3$, $-13.5 \pm 0.4$, and $4.1 \pm 1.3$ \kms, respectively. According to the classifications of \citet{Bensby:2003}, this gives KELT-16 a 99.4 \% probability of belonging to the thin disk population of the Galaxy. Furthermore, these relatively low peculiar velocities are consistent with the best-fit isochronal age of $3.1 \pm 0.3$ Gyr (\S\ref{sec:HRD}).

\subsection{Transit Timing Variation Analysis}
\label{sec:TTVs}

Variations in the observed periodicity of a planet's transit, or transit timing variations (TTVs), can be caused by the gravitational perturbations of other planets in the same system, and thus serve as a means for discovering those planets \citep{Holman:2005, Agol:2005}. Models predict that TTVs will usually be largest for planets in orbital mean motion resonances (MMRs; e.g., \citealt{Agol:2005}). However, until very recently, TTVs had not been detected for hot Jupiters. This could be because hot Jupiters' co-planets are generally of low mass or in orbits unlikely to cause MMRs, or because hot Jupiters eject most of their co-planets during their inward migration (e.g., \citealt{Mustill:2015}). However, \citet{Becker:2015} recently detected a TTV of amplitude 38 seconds in WASP-47b driven by a Neptune-mass planet in a nearby superior orbit (and possibly also to a lesser degree by a super-Earth in an inferior orbit). This demonstrates that multi-planet systems can induce TTVs in hot Jupiters, and that these TTV amplitudes can be large enough to detect from the ground.

However, for KELT-16b, the uncertainties in the mid-transit times, $T_\textrm{C}$, of the global fit to the individual KELT-FUN light curves range from $42 \leq \sigma_{T_\textrm{C}} \leq 150$ s, with only about half of them having values of $\sim 1$ minute or less (Tables \ref{tab:KELT-16b_2} or \ref{tab:TTVs}). Even taking only this best half, KELT-16b would need to have a TTV signal of $\sim 180$ s to be detected at the $3 \sigma$ level by KELT-FUN. Such a high TTV signal is expected to be rare for a hot Jupiter, but is theoretically possible even for a low-mass perturbing planet for special combinations of that planet's eccentricity and inclination \citep{Payne:2010}.


We thus check KELT-16b for TTV signals. To do so, we compute the O-C residuals between the global fit mid-transit times for the 19 KELT-FUN light curves, $T_\textrm{C}$, and the global fit ephemeris, $T_\textrm{C} = 2457247.24791\pm0.00019$ \bjdtdb and $P = 0.9689951 \pm 0.0000024$ d (Table \ref{tab:KELT-16b_1}). Note that, during the global fit, the $T_\textrm{C}$ for each light curve was fit individually and independently of the others -- i.e., $P$ was fit to the $T_\textrm{C}$'s and not the other way around -- so any TTVs that may exist would not have been altered by the global fit itself. 

\begin{figure}[!ht]
\centering\epsfig{file=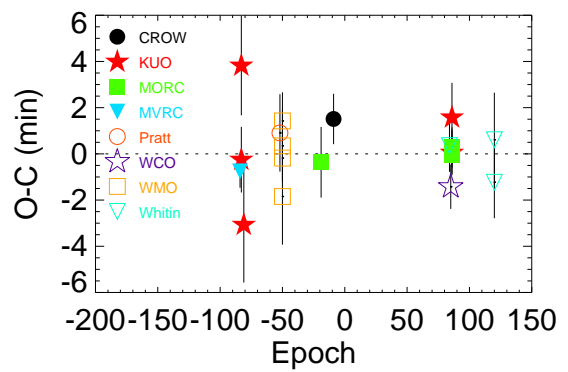,clip=,width=0.99\linewidth}
\caption{\footnotesize Transit time residuals for KELT-16b relative to the global fit ephemeris (flat dashed line). The data are listed in Table \ref{tab:TTVs}.}
\label{fig:TTVs}
\end{figure}

The results are listed in Table \ref{tab:TTVs} and plotted in Figure \ref{fig:TTVs}. None of the data points are found to have a statistically significant deviation from the global fit ephemeris (the flat dashed line): the average O-C is 60 s or $0.7 \sigma$, with the largest single outlier being 230 s or $1.8 \sigma$. We thus do not believe these outliers are astrophysical in origin and attribute them to observational systematics as described by \citet{Carter:2009}. We carefully ensured that all follow-up observations were correctly converted to BJD$_{\rm TBD}$ \citep{Eastman:2010} and that each observatory clock was synchronized with a standard clock to sub-millisecond accuracy (\S\ref{sec:Photom}) such that the limiting uncertainty for each $T_\textrm{C}$ is the global fit (and therefore the quality of the photometric data) itself, rather than any systematic differences in timekeeping by observers.  

\begin{table}
 \centering
 \caption{Transit times for KELT-16\MakeLowercase{b}.}
 \label{tab:TTVs}
 \begin{tabular}{r@{\hspace{12pt}} l r r r c}
    \hline
    \hline
    \multicolumn{1}{c}{Epoch} & \multicolumn{1}{c}{$T_\textrm{C}$} 	& \multicolumn{1}{l}{$\sigma_{T_\textrm{C}}$} 	& \multicolumn{1}{c}{O-C} &  \multicolumn{1}{c}{O-C} 			& Telescope \\
	    & \multicolumn{1}{c}{(\bjdtdb)} 	& \multicolumn{1}{c}{(s)}			& \multicolumn{1}{c}{(s)} &  \multicolumn{1}{c}{($\sigma_{T_\textrm{C}}$)} 	& \\
    \hline
 -84 & 2457165.85179 &   42  &  -45  & -1.1 & MVRC \\
 -83 & 2457166.82114 &   85  &  -15  & -0.2 & KUO \\
 -83 & 2457166.8240  &  130  &  230  &  1.8 & KUO \\
 -81 & 2457168.7572  &  150  & -180  & -1.2 & KUO \\
 -52 & 2457196.8608  &  100  &   50  &  0.5 & Pratt \\
 -50 & 2457198.7984  &   90  &   20  &  0.2 & WMO \\
 -50 & 2457198.79803 &   56  &  -11  & -0.2 & WMO \\
 -50 & 2457198.79914 &   74  &   85  &  1.1 & WMO \\
 -50 & 2457198.7969  &  120  & -110  & -0.9 & WMO \\
 -19 & 2457228.8368  &   90  &  -20  & -0.2 & MORC \\
  -9 & 2457238.52801 &   65  &   91  &  1.4 & CROW \\
  84 & 2457328.64376 &   68  &   21  &  0.3 & Whitin \\
  85 & 2457329.61152 &   57  &  -85  & -1.5 & WCO \\
  86 & 2457330.58170 &   52  &   17  &  0.3 & MORC \\
  86 & 2457330.58146 &   48  &   -3  & -0.1 & MORC \\
  86 & 2457330.58154 &   58  &    3  &  0.1 & KUO \\
  86 & 2457330.5826  &   90  &   90  &  1.0 & KUO \\
 120 & 2457363.5265  &   90  &  -70  & -0.8 & Whitin \\
 120 & 2457363.5278  &  120  &   40  &  0.3 & Whitin \\
    \hline
    \hline
 \end{tabular}
\end{table}

\subsection{Tidal Evolution and Irradiation History}
\label{sec:Irradiation}

Following \citet{Penev:2014}, we model the orbital evolution of KELT-16b due to the dissipation of the tides raised by the planet on the the host star under the assumption of a constant phase lag. The starting configuration of the system was tuned to reproduce the presently observed system parameters (Table \ref{tab:KELT-16b_1}) at the assumed system age of 3.1 Gyr (\S\ref{sec:HRD}). The evolution includes the effects of the changing stellar radius and luminosity following the YY circular stellar model with mass and metallicity as given in Table \ref{tab:KELT-16b_1}, but neglects the effects of the stellar rotation, assuming that the star is always rotating sub-synchronously relative to the orbit.

\begin{figure}
\includegraphics[width=1\linewidth]{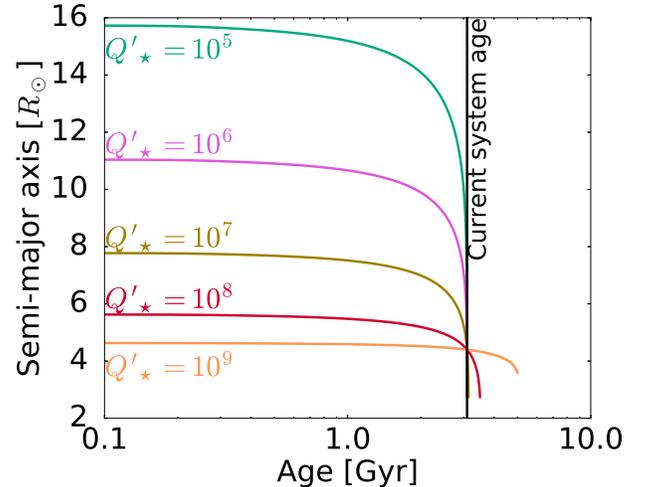}
   \vspace{-.2in}
   \caption{\footnotesize
   The orbital semi-major axis history of KELT-16b modeled for a range of stellar tidal quality factors, $Q_*'$, where $Q_*'^{-1}$ is the product of the tidal phase lag and the Love number. The black vertical line marks the current system age of 3.1 Gyr.}
   \label{fig:Tides}
\end{figure}

\begin{figure}   
\includegraphics[width=1\linewidth]{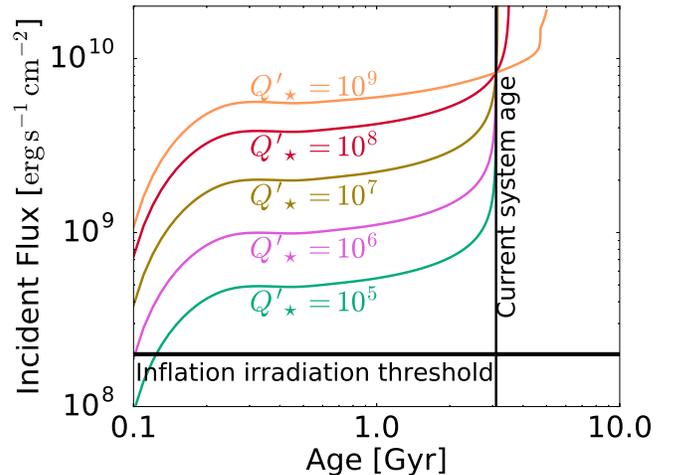}
   \vspace{-.2in}
   \caption{\footnotesize
   The irradiation history of KELT-16b modeled for a range of stellar tidal quality factors, $Q_*'$, where $Q_*'^{-1}$ is the product of the tidal phase lag and the Love number. The black vertical line marks the current system age of 3.1 Gyr while the black horizontal line marks the inflation irradiation threshold of $\approx 2 \times 10^8$ ergs s$^{-1}$ cm$^{-2}$  \citep{Demory:2011}.
   }
   \label{fig:Irradiation}
\end{figure}

Orbital and stellar irradiation evolutions are shown in Figures \ref{fig:Tides} and \ref{fig:Irradiation} for a range of stellar tidal quality factors ($Q_*' = 10^5, 10^6, 10^7, 10^8, \textrm{and}\, 10^9$), where $Q_*'^{-1}$ is the product of the tidal phase lag and the Love number. We find that the insolation received by the planet is well above the empirical inflation irradiation threshold ($\approx 2 \times 10^8$ ergs s$^{-1}$ cm$^{-2}$; \citealt{Demory:2011}) for the entire main-sequence existence of the star, except in the very early stages of stellar evolution for the case $Q_*' = 10^5$ (Figure \ref{fig:Irradiation}).

We consider a wide range of $Q_*'$ because of the wide range of proposed mechanisms for tidal dissipation in current theoretical models and the conflicting observational constraints backing those models, especially for stars that may have surface convective zones (see the review by \citealt{Ogilvie:2014} and references therein). Furthermore, because the dependence on stellar mass and tidal frequency is different for the different proposed mechanisms, we make the simplifying assumption that $Q_*'$ remains constant over the life of the star. However, with multi-year baselines, it may be possible in the future to empirically constrain the lower limit on $Q_*'$ for KELT-16 via precise measurements of the orbital period time decay (cf. \citealt{Hoyer:2016}).

Finally, note that this model does not account in any way for the larger-distance Type II or scattering-induced migration that KELT-16b and other hot Jupiters likely undergo. It considers only the close-in migration due to tidal friction alone.

%
%
%

\section{False-Positive Analysis}
\label{sec:False-Positives}


Our usual first step in vetting false-positive planet candidates is to obtain additional time-series photometry using KELT-FUN (\S \ref{sec:Photom}). Compared to the KELT-North telescope (\S\ref{sec:Discovery}), these telescopes can provide seeing-limited resolution, lower photometric noise, higher time cadences, and multi-band photometry (Table \ref{tab:Photom}). This quickly vets artifacts, variables and EBs based on their light curve profiles. Grazing and blended EBs can still exhibit shallow, flat-bottomed eclipses, but can be vetted by looking for wavelength-dependent depths. In the 19 KELT-FUN full transit light curves of KELT-16b spanning the $B$ through $z'$ filters, we find the depth to be achromatic and the transit profile to be well-fit by a planetary model (\S\ref{sec:Photom}). 

The usual second step is to obtain spectroscopic observations, which provide several additional independent checks. We visually inspected the 20 spectroscopic follow-up measurements of KELT-16 by TRES (\S\ref{sec:Spectra}) and find no double spectral lines due to a blended stellar companion. We find an RV signal consistent with a planetary-mass companion and find no correlation between the bisector spans and measured RVs \citep{Buchave:2010}. Furthermore, the stellar surface gravity derived from the spectra, log($g_\star$) = 4.03 $\pm$ 0.11 (\S\ref{sec:SpecParams}), is roughly consistent with that derived from the transit via the global fit,  $4.253_{-0.036}^{+0.031}$ (\S\ref{sec:GlobalFit}).

As a third step, we obtain high contrast, high spatial resolution AO imaging to search for nearby companions that may be blended in the follow-up photometry and spectroscopy. As described in \S\ref{sec:AO}, a companion is detected. 
%
We therefore 
consider the possibility that the observed transits are due to a body orbiting the companion star rather than the primary star. The observed transits have a best-fit depth of $\delta = 11.46_{-0.25}^{+0.29}$ mmag and are achromatic to within the $\sim 1$ mmag rms uncertainty of the best individual KELT-FUN light curves (\S\ref{sec:Photom} and \S\ref{sec:GlobalFit}). Thus, we can rule out the circum-companion scenario if the limits on the blended transit depth in the filters observed by KELT-FUN are either (a) shallower than this depth (in \textit{any} of the filters) and/or (b) chromatic (have a depth difference in any two of the filters) at levels detectable above the KELT-FUN rms uncertainty.

To compute blended depths in multiple filters it is necessary to know the color of the companion star, which has not been directly measured, but can be modeled if the distance to the companion is assumed. Although the companion is likely bound to the KELT-16 primary and thus at a distance of $\sim 399 \pm 19$ pc, the circum-companion scenario does not require boundedness, so, to cover all bases, we consider a range of distances to the companion. KELT-16 has a Galactic latitude of $b = -8\fdg94227$ and is likely in the Galactic thin disk (\S\ref{sec:UVW}). Assuming a thin disk scale height of $z_d \sim 300$ pc \citep{Juric:2008}, the companion is thus unlikely to be more than $\sim 2000$ pc distant. Combining this distance with the companion's observed apparent magnitude of $K_\textrm{S} = 15.0$ yields an absolute magnitude of $M_{K_\textrm{S}} \sim 3.5$, which, based on the color-temperature sequence of \citet{Pecaut:2013},\footnote{\url{http://www.pas.rochester.edu/~emamajek/}}
limits the companion to $V > 16.7$. But, at this faintness, even the \textit{complete occultation} of the companion by a dark circum-companion object would only produce a $\delta_V = 10.0$ mmag transit depth. The situation is worse in the $B$-band, where total occultation would only produce a $\delta_B = 8.8$ mmag transit depth. And the situation is worse still if the companion is less distant, the occultation is incomplete, or the circum-companion body is not dark. In summary, the circum-companion scenario is unable to produce transits as deep as those observed in the bluer optical filters \textit{regardless of the distance to or physical nature of either the companion or circum-companion body}. We can thus rule out completely the circum-companion false-positive scenario based on the transit depth alone.

Note that chromaticity could independently be used to rule out the circum-companion scenario in some cases, but not all. Transit depth differences between the $B$ and $z'$ filters would be as large as $\delta_B - \delta_{z'} \sim 3.5$ mmag when the companion is very distant, the circum-companion body dark, and the occultation complete, but could be shallower than 1 mmag -- and thus unlikely to be detected by KELT-FUN -- in other cases.
%

Finally, note also that the $V > 16.7$ limit on the companion brightness means it is
too faint for its spectrum to distort the velocities of the primary by the observed amount, and thus could not be responsible for the RV signal observed by TRES.
 
Several lines of evidence thus exist to rule out each possible false-positive scenario. We conclude that all of the available data are best explained by a planetary companion orbiting KELT-16.


\section{Discussion}
\label{sec:Discussion}

\subsection{Comparative Planetology}
\label{sec:ComparativePlanetology}

\begin{figure}
\includegraphics[width=0.85\linewidth, angle = -90]{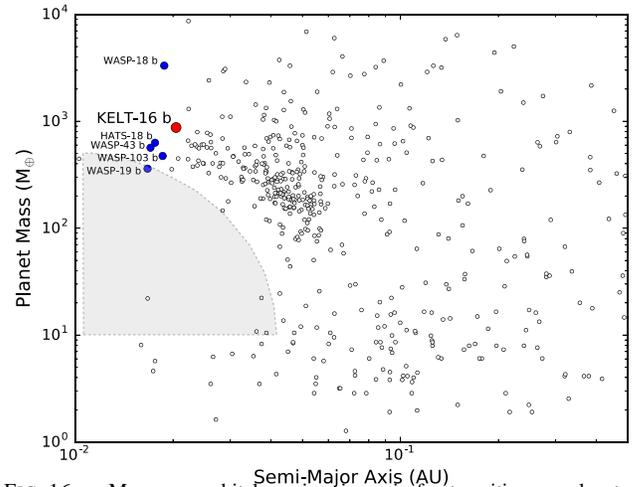}
   \vspace{-.2in}
   \caption{\footnotesize 
   Mass vs. orbital semi-major axis for transiting exoplanets. KELT-16b and the five other $P < 1$ d giant planets are indicated by the larger red and blue data points, respectively. The irradiation-driven evaporation zone for a 30 $M_\Earth$ core planet from the model of \citet{Kurokawa:2014} is approximately depicted by the shaded region. Exoplanet data were obtained from the NASA Exoplanet Archive on July 21, 2016.}
   \label{fig:M-a}
\end{figure}

\begin{figure}   
\includegraphics[width=1.08\linewidth]{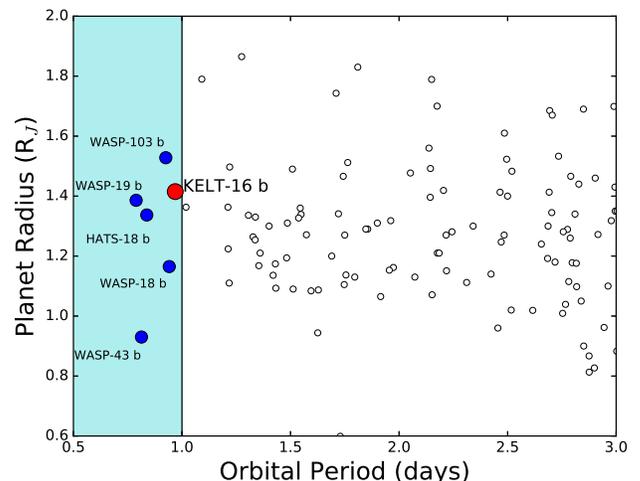}
   \vspace{-.2in}
   \caption{\footnotesize
   Planet radius vs. orbital period for transiting exoplanets. KELT-16b and the five other $P < 1$ d giant planets are indicated by the larger red and blue data points, respectively. Exoplanet data were obtained from the NASA Exoplanet Archive on July 21, 2016.}
   \label{fig:R-P}
\end{figure}

KELT-16b is in a select group of ultra-short period inflated hot Jupiters transiting relatively bright host stars. There are only five others with $< 1$ d periods, WASP-18b, -19b, -43b, -103b, and HATS-18b (\citealt{Hellier:2009, Hebb:2010, Hellier:2011, Gillon:2014}; and \citealt{Penev:2016}, respectively), and only one of these, WASP-18b, is more massive than KELT-16b (Figures \ref{fig:M-a} and \ref{fig:R-P}). Such planets are subject to extreme irradiation and strong tidal forces, making it plausible that long-term follow-up of these systems could reveal evidence of atmospheric evolution or orbital decay (e.g., \citealt{Hoyer:2016}). Furthermore, comparative studies of such planets could inform planet formation and evolution scenarios, in particular how significant a role planet mass plays in the rate of atmospheric or dynamic evolution.

%
%
%

\subsection{Prospects for Atmospheric Characterization}
\label{sec:Atmosphere}

Because of its high irradiation and large size, KELT-16b has a higher infrared flux than most planets. It is thus expected to have larger phase curve amplitudes and secondary eclipse depths -- and thus higher SNR -- in the HST and Spitzer infrared passbands.
This makes KELT-16b a promising target for tackling a number of open questions in the field right now regarding exoplanetary atmospheres, such as the presence and cause of temperature-pressure inversions, the identification of upper atmosphere absorbers at hot temperatures, and day-to-night heat transfer. 

Due to its high mass, we find KELT-16b to have a relatively small atmospheric scale height of $H = k T_\textrm{eq} / \mu g_\textrm{P} = 267 \pm 15$ km, where we assume a hydrogen-dominated atmosphere with a mean molecular weight of $\mu = 2$ following \citet{Winn:2010b}, $T_\textrm{eq}$ and $g_\textrm{P}$ are taken from our global fit (Table \ref{tab:KELT-16b_1}), and $k$ is the Boltzmann constant. Thus, measuring the wavelength-dependent radius during transit to produce a transmission spectrum is not as compelling with current telescopes, but KELT-16b will be a good candidate for such observations with the larger aperture JWST.
Therefore, a comprehensive study of KELT-16b in both transit and eclipse will be possible in the near future.


Of particular interest, KELT-16b is a higher mass, higher temperature analog of the well-studied planet WASP-43b, providing an opportunity for several key differential measurements. For example, due to its high level of irradiation, KELT-16b falls into the ``pM class'' of planets that are warm enough to have opacity due to TiO and VO gasses, while WASP-43b may belong to the cooler ``pL class'' of planets in which the TiO and VO have condensed and rained out 
\citep{Fortney:2008}.
pM planets such as KELT-16b are expected to absorb incident stellar flux in the stratosphere, leading to temperature-pressure inversions and large day-to-night temperature swings (hotter days and cooler nights). pL planets, on the other hand, are expected to absorb incident flux deeper in the atmosphere where atmospheric dynamics will more readily redistribute absorbed energy, leading to less drastic day-to-night temperature swings (cooler days and hotter nights). If KELT-16b's temperature cools enough in going from the dayside to the nightside, TiO and VO could rain out at the terminator and its temperature-pressure inversion disappear on the night side (e.g., \citealt{Burrows:1999, Lee:2015}).



\begin{figure}
\includegraphics[width=1\linewidth]{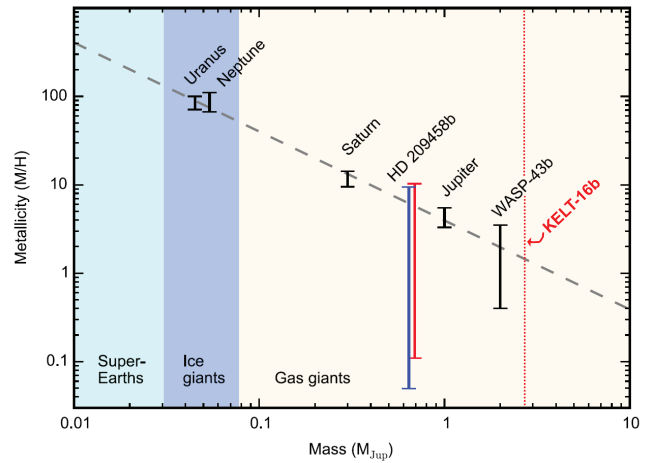}
\caption{\footnotesize
Atmospheric metal abundances as a function of planet mass for the Solar System giant planets and WASP-43b. A measurement of KELT-16b's metallicity could test this relation out to nearly 3 $M_\textrm{J}$ (dotted red line). The solid red and blue bars show two different measurements for HD 209458b; see \citet{Line:2016} for details. Reproduced from \citet{Kreidberg:2014} Figure 4 and \citet{Line:2016} Figure 10 with permission. }
\label{fig:Kreidberg2014_Fig4} 
\end{figure}

Finally, KELT-16b presents an opportunity to study the trend observed in Solar System giants of decreasing metallicity with increasing planet mass. This relationship was extended to higher mass by WASP-43b \citep{Kreidberg:2014} and could be tested out to nearly 3 $M_\textrm{J}$ with KELT-16b (Figure \ref{fig:Kreidberg2014_Fig4}).

\subsection{Orbital Dynamics}
\label{sec:Dynamics}

Because most formation theories require that giant planets form beyond the snow line, KELT-16b may have once orbited at a distance of $\gtrsim 5$ AU with a period of $\sim 10$ yr or more \citep{Kennedy:2008, Garaud:2007} before migrating inward due to Type II or scattering-driven migration (\S\ref{sec:Intro}). 
The likely existence of a bound, widely separated stellar companion in the KELT-16 system (\S\ref{sec:Companion}) presents the possibility that the Kozai-Lidov (KL) mechanism influenced this migration \citep{Kozai:1962,Lidov:1962}. If KELT-16b formed beyond the snow line, its KL timescale could have been as short as $T_\textrm{KL} \sim 2.5$ Myr (e.g., \citealt{Kiseleva:1998}), assuming the current minimum circular orbital period of the stellar companion (Table \ref{tab:Companion}). If the stellar companion orbit is eccentric, then the minimum KL timescale could have been far less than this
Future observations of the Rossiter-McLaughlin (RM) effect could help provide evidence for past KL librations by measuring system misalignment. It is also noteworthy that KELT-16 sits nearly exactly at the $T_\textrm{eff}$ = 6250 K boundary for misalignment suggested by \citet{Winn:2010a}. 

Once KELT-16b had migrated to within a few tens of stellar radii, its orbital dynamics became dominated by tidal forces. Based on the modeling of these forces, KELT-16b began a runaway in-spiral by the age of $\sim$ 1 Gyr (\S\ref{sec:Irradiation}). It is currently orbiting at a radius of $a = 3.373 R_\star = 4.382 R_\odot$, just $\sim$ 1.7 times the Roche limit of $a_\textrm{R} = 2.009 \pm 0.045 R_\star = 3.373 \pm 0.022 R_\odot$. This is within the 2 $a_\textrm{R}$ radius at which planets scattered inward and having eccentric orbits are expected to tidally circularize \citep{Matsumura:2010}, consistent with the circular orbital solutions of the global fit (\S \ref{sec:GlobalFit}). 

At such a close distance to its star, it is worth exploring whether KELT-16b may be in danger of either evaporation due to extreme irradiation or disintegration due to extreme tidal forces.  
KELT-16b's current x-ray and extreme UV flux- ($F_\textrm{XUV}$)-driven atmospheric mass loss rate is expected to be $\dot{M} < 4.0 \times 10^{-13} M_\textrm{J}$ yr$^{-1}$, as calculated via equations 6 and 7 of \citet{Salz:2015} using the KELT-16 system parameters derived in this work and assuming that KELT-16 has the same upper limit on $F_\textrm{XUV}$ as measured by \citet{Salz:2015} for the similar star WASP-18. 
Even if this rate persisted for the life of KELT-16b (untrue since the irradiation evolves; Figure \ref{fig:Irradiation}), the total mass loss over the life of KELT-16b to date would be $\sim 0.0012 M_\textrm{J}$, or $\sim 0.045$ \% of the planet's total mass. Thus, although KELT-16b is highly irradiated, its high overall mass limits the evaporative mass loss rate and keeps it safely above the evaporation zone as defined by the models of Kurokawa and Nakamoto (2014; approximately depicted by the shaded region in Figure \ref{fig:M-a}). 

Nevertheless, KELT-16b will soon be tidally shredded as it continues a runaway in-spiral towards its star. Based on the orbital evolution model of \S\ref{sec:Irradiation}, KELT-16b could cross the Roche limit in a mere $\sim$ 550,000 years if the stellar tidal quality factor is $Q_*' = 10^5$, or in as long as $\sim$ 2.5 Gyr from now -- roughly doubling the planet's current age and lasting nearly as long as its star -- if $Q_*' = 10^9$.


\subsection{Future observations}
\label{sec:FutureObs}


KELT-16b's ultra-short period offers a number of practical advantages for future observations. First, it means that observing a full orbital phase requires less telescope time -- an important consideration for oversubscribed telescopes such as HST, Spitzer, the soon-to-be-launched JWST, and others. The short period also makes it easier for most observatories to cover a higher fraction of the full phase curve in a single continuous observation. And when more telescope time is available, a short period allows for the more rapid observation and phase-folded stacking of multiple orbits to achieve a high SNR. The other $P < 1$ d giant planets have been heavily observed in part for these reasons. However KELT-16b has the additional advantage of its period being very close to (just 45 minutes shy of) a sidereal day, making it unusually easy to also collect multiple consecutive transit or secondary eclipse observations from the ground when it is ``in phase'' for a particular observatory.

In addition to its short period, KELT-16b has the eighth highest incident flux, and thus equilibrium temperature ($T_\textrm{eq} \approx 2400$ K), amongst giant planets transiting bright stars,
giving it a large phase curve amplitude and secondary eclipse signal. Finally, its relatively bright ($V = 11.7$) host offers moderately high photometric precision.

KELT-16b thus has all of the characteristics to become presently one of the top observational targets amongst exoplanets.


\section{Acknowledgements}

We thank an anonymous referee for thoughtfully reviewing this manuscript and offering several helpful suggestions. 

T.E.O. acknowledges a sabbatical award from Westminster College. B.S.G. and D.J.S. acknowledge support from NSF CAREER Grant AST-1056524. D.A. acknowledges support from an appointment to the NASA Postdoctoral Program at Goddard Space Flight Center, administered by the Universities Space Research Association (USRA) through a contract with NASA. K.K.M. acknowledges the purchase of SDSS filters for Whitin Observatory by the Theodore Dunham, Jr., Grant of the Fund for Astronomical Research. The AO data in this work were obtained at the W.M. Keck Observatory, which was financed by the W.M. Keck Foundation and is operated as a scientific partnership between the California Institute of Technology, the University of California, and NASA.

In addition, this research has made use of the following services and databases: The Smithsonian Astrophysical Observatory/NASA Astrophysics Data System (SAO/NASA ADS); Vizier \citep{Ochsenbein:2000}; The SIMBAD Astronomical Database \citep{Wenger:2000}; The Spanish Virtual Observatory (SVO) Filter Profile Service (\url{http://svo2.cab.inta-csic.es/svo/theory/fps3/index.php?mode=browse});
and the American Association of Variable Star Observers (AAVSO) Photometric All-Sky Survey (APASS), whose funding is provided by the Robert Martin Ayers Sciences Fund and the AAVSO Endowment (\url{https://www.aavso.org/aavso-photometric-all-sky-survey-data-release-1}).

\bibliographystyle{apj}

\bibliography{KELT-16b}

\end{document}